\documentclass{article}

\usepackage{arxiv}
\usepackage{amsthm, amsmath, amsbsy, amssymb, booktabs, epsfig, epsf}
\usepackage{multicol, multirow, psfrag, booktabs, graphicx, rotating,subfigure}
\usepackage[authoryear]{natbib}
\usepackage[colorlinks, citecolor = blue, urlcolor = blue]{hyperref}
\usepackage{tikz}
\usepackage[right]{lineno}
\usepackage{tabularx}
\usepackage{subcaption}

\title{The Analysis of Criminal Recidivism: A Hierarchical Model-Based Approach for the Analysis of Zero-Inflated, Spatially Correlated recurrent events Data}

\author{
  Alisson C. C. Silva \\
  Department of Statistics\\
  Universidade Federal de Minas Gerais\\
  Av. Presidente Ant\^{o}nio Carlos 6627 \\
  Pampulha Belo Horizonte, Minas Gerais \\
  31270-901, Brazil\\
  \texttt{alisson.ccs2@gmail.com} \\
  \And
  Fábio N. Demarqui \\
  Department of Statistics\\
  Universidade Federal de Minas Gerais\\
  Av. Presidente Ant\^{o}nio Carlos 6627 \\
  Pampulha Belo Horizonte, Minas Gerais \\
  31270-901, Brazil\\
  \texttt{fndemarqui@est.ufmg.br} \\
  \And
  Bráulio F. Silva \\
  Department of Sociology \\ 
  Universidade Federal de Minas Gerais\\
  Av. Presidente Ant\^{o}nio Carlos 6627 \\
  Pampulha Belo Horizonte, Minas Gerais \\
  31270-901, Brazil\\
  \texttt{braulio.fas@gmail.com} \\
  \And
  Marcos O. Prates \\
  Department of Statistics\\
  Universidade Federal de Minas Gerais\\
  Av. Presidente Ant\^{o}nio Carlos 6627 \\
  Pampulha Belo Horizonte, Minas Gerais \\
  31270-901, Brazil\\
  \texttt{marcosop@est.ufmg.br} \\
}

\begin{document}
\maketitle

\begin{abstract}
The life course perspective in criminology has become prominent last years, offering valuable insights into various patterns of criminal offending and pathways. Noticeably, the study of criminal trajectories aims to understand the beginning, persistence and desistence in crime, providing intriguing explanations about these moments in life. The Criminal Career Approach is a fundamental framework in this field, recognizing that individuals start committing crimes at specific times, become involved in different types of criminal activities, and eventually abandon this career. 
Central to this analysis is the identification of patterns in the frequency of criminal victimization and recidivism, along with the factors that contribute to them. Specifically, this work introduces a new class of models that overcome limitations in traditional methods used to analyze criminal recidivism. These models are designed for recurrent events data characterized by excess of zeros and spatial correlation. They extend the Non-Homogeneous Poisson Process, incorporating the structure of the Intrinsic Conditional Auto-Regressive model through random effects, enabling the analysis of associations among individuals within the same spatial stratum. To deal with the excess of zeros in the data, the structure of the zero-inflated Poisson mixed model was incorporated. In addition to parametric models following the Power Law process for baseline intensity functions, we propose flexible semi-parametric versions approximating the intensity function using Bernstein Polynomials. The Bayesian approach offers advantages such as incorporating external evidence and modeling specific correlations between random effects and observed data. The performance of these models was evaluated in a simulation study with various scenarios, and we applied them to analyze criminal recidivism data in the Metropolitan Region of Belo Horizonte, Brazil. The results provide a detailed analysis of high-risk areas for recurrent crimes and the behavior of recidivism rates over time. This research significantly enhances our understanding of criminal trajectories, paving the way for more effective strategies in combating criminal recidivism.
\end{abstract}


\section{Introduction}

The life course perspective in criminology has become prominent last years, offering valuable insights into various patterns of criminal offending and pathways \citep{petersilia1980criminal, blumstein1986criminal, Piquero_Farrington_Blumstein_2007}. The \emph{Criminal Career Approach} is one of the main structures in criminology, recognizing that individuals start their criminal activity at a certain point, engage in crime, commit various crimes, and eventually give up their career \citep{blumstein1986criminal, piquero2003criminal, delisi2005career}. The identification of patterns of constancy and change in the frequency with which individuals engage in criminal acts, as well as the identification of factors that impact this frequency, is one of the central points of the analysis of criminal careers \citep{paternoster1997multiple, blumstein2010linking}. Nonetheless, the diversity of approaches to define the term \emph{criminal recidivism} poses challenges to formulating indicators of recidivism.

In a broader definition, criminal recidivism can be described as the phenomenon of repeating criminal acts and delinquent trajectories established throughout the course of life. From a legal perspective, the Brazilian Criminal Code defines a recidivism offender as an individual who, after a final and unappealable decision regarding a previous crime, commits a new crime within 5 years. \citet{ribeiro2022} analyzed 144 texts related to criminal recidivism in Brazil and identified 5 classifications for the type of recidivism. Among them, prison recidivism and police recidivism stand out. Prison recidivism characterizes an individual as a recidivism offender if they have served a term of imprisonment and return to prison, regardless of a new conviction. Police recidivism characterizes an individual as a recidivism offender if they commit more than one crime recorded by the police or the Judiciary.
Studies focusing on prison administration and public policies for prisoners and individuals with prior involvement in the justice system generally explore prison recidivism, while research focused on public safety usually prioritizes police recidivism. These studies aim to support the formulation of public policies aimed at reducing crime, controlling risk factors, facilitating the social reintegration of former inmates, and assessing the risk of committing new crimes.

In the Brazilian context, studies such as those conducted by \citet{ipea2015reincidencia} and \citet{gappe2022reincidencia} evaluate recidivism rates in Brazilian states, defined as the proportion of individuals released from the prison system who return to commit new crimes. Additionally, \citet{sapori2017fatores} study the association between the probability of recidivism and the sociodemographic and criminal profiles of individuals released from the prison system in the state of Minas Gerais, utilizing logistic regression models for analysis.
The use of statistical models as a tool for predicting and assessing the risk of recidivism is a widespread practice in several countries. One example is RisCanv \citep{andres2018riscanvi}, a risk assessment tool adopted in the prison system of Catalonia, Spain. Employing ROC curves and survival analysis methodologies, this tool gauges the likelihood of an individual engaging in subsequent recidivism based on data collected from inmates and individuals with prior involvement in the justice system.
The work of \cite{Heijden2013} includes a comparative analysis between the performance of classical models and machine learning models used to predict criminal recidivism. Another approach employs the use of survival analysis to analyze the time from an offender's conviction, or release from prison, to the first conviction. Among other articles, the use of survival models to analyze criminal recidivism is discussed in \cite{Stollmark1974} and \cite{Copas2008}. In general terms, such studies focus on analyzing the recidivism rate, calculating the probability of recidivism and analyzing the time until the first recidivism occurs. However, such approaches do not allow for a broader analysis that considers the individual's trajectory and the frequency of committing crimes.

The literature offers a range of methods suitable for analyzing recurrent events by modeling the time between events or the number of events experienced by individuals \citep{rigdon2000statistical,cook2007statistical, kleinbaum2012survival, meeker2022statistical}. In the context of criminal recidivism, applying these methods would enhance the analysis of multiple recidivism events over time, allowing for the assessment of fluctuations in the frequency of these occurrences.

This study introduces an innovative methodology designed to provide significant insights for addressing pertinent questions regarding criminal recidivism. The introduced methods aim to address inquiries such as: Is there a discernible trend regarding the frequency of recidivism? What factors are linked to the frequency of recidivism? What is the probability of recidivism? Does geographic space exert an influence on the recidivism pattern? The questions pose challenges to traditional methods used for analyzing recurrent events data, as current approaches cannot address issues of great importance for researchers. Specifically, these issues include the absence of a framework that enables the assessment of the impact of geographical space on recidivism behavior, as well as the inability to effectively handle the issue of excess of zeros related to the frequency of recurrences. In this context, a series of novel models will be introduced for analyzing criminal recidivism data, effectively overcoming the limitations of conventional approaches. These models are applicable to the analysis of various types of criminal recidivism and provide more comprehensive and sophisticated metrics. Utilizing counting methods, the proposed models integrate a framework capable of handling zero-inflated data and incorporate a random effects structure that enables the evaluation of spatial effects on both the frequency and probability of recidivism.  
Parametric and semiparametric versions of these models will be introduced, with the baseline intensity function approximated using Bernstein polynomials \citep{bernstein1912demo}. The adoption of the Bayesian approach in this study brings several advantages, including the formal integration of external empirical evidence and the capability to induce specific correlations among random effects and observed data. It worth mentioning that the proposed models have many of the conventional ones as special cases.

The remainder of this article follows the following structure: The Database section presents the dataset made available for the analysis of criminal recidivism in the Metropolitan Region of Belo Horizonte. The Literture Review section discusses traditional models for analyzing recurrent event data. Next, in the Methodology section, the contribution of this article is detailed, presenting a set of models for the analysis of recurrent event data, spatially correlated and zero-inflated. The Inference section discusses the necessary elements for the use of the Bayesian approach, as well as the information criteria used in this study. The Simulation Study and Application sections address the study of simulated data and the analysis of real data, respectively. Finally, in the Conclusion section, the final considerations related to this study are presented.

\hypertarget{The_dataset}{%
\section{Database}\label{The_dataset}}
The dataset used for the analysis of criminal recidivism comprises information extracted from the Records of Social Defense Events (RSDE), spanning from January 2013 to December 2019. These records pertain to incidents of aggravated assault\footnote{The FBI’s Uniform Crime Reporting (UCR) Program defines aggravated assault as an unlawful attack by one person upon another for the purpose of inflicting severe or aggravated bodily injury.} documented in the Municipality of Belo Horizonte, Minas Gerais, as well as in four neighboring municipalities: Betim, Contagem, Ribeirão das Neves, and Santa Luzia. The RSDE dataset includes information such as the number of individuals involved, an individual identifier\footnote{The individual identifier code was treated with a randomization mechanism, thus protecting each individual's information in compliance with the Brazilian General Data Protection Law.} code, genders, parents' names, and their classification based on involvement (victim, offender, principal, and suspect), as well as the date and time of the incident. Additionally, geographic coordinates (longitude and latitude) are provided to indicate the location of each crime. The dataset encompasses data from 23,700 individuals, categorized as offenders, principals, or suspects. Notably, only $22.74\%$ of the individuals involved in the overall records are female.

\begin{figure}[h!]
\centering
\begin{minipage}{0.48\linewidth}
  \centering
  \includegraphics[width=\linewidth]{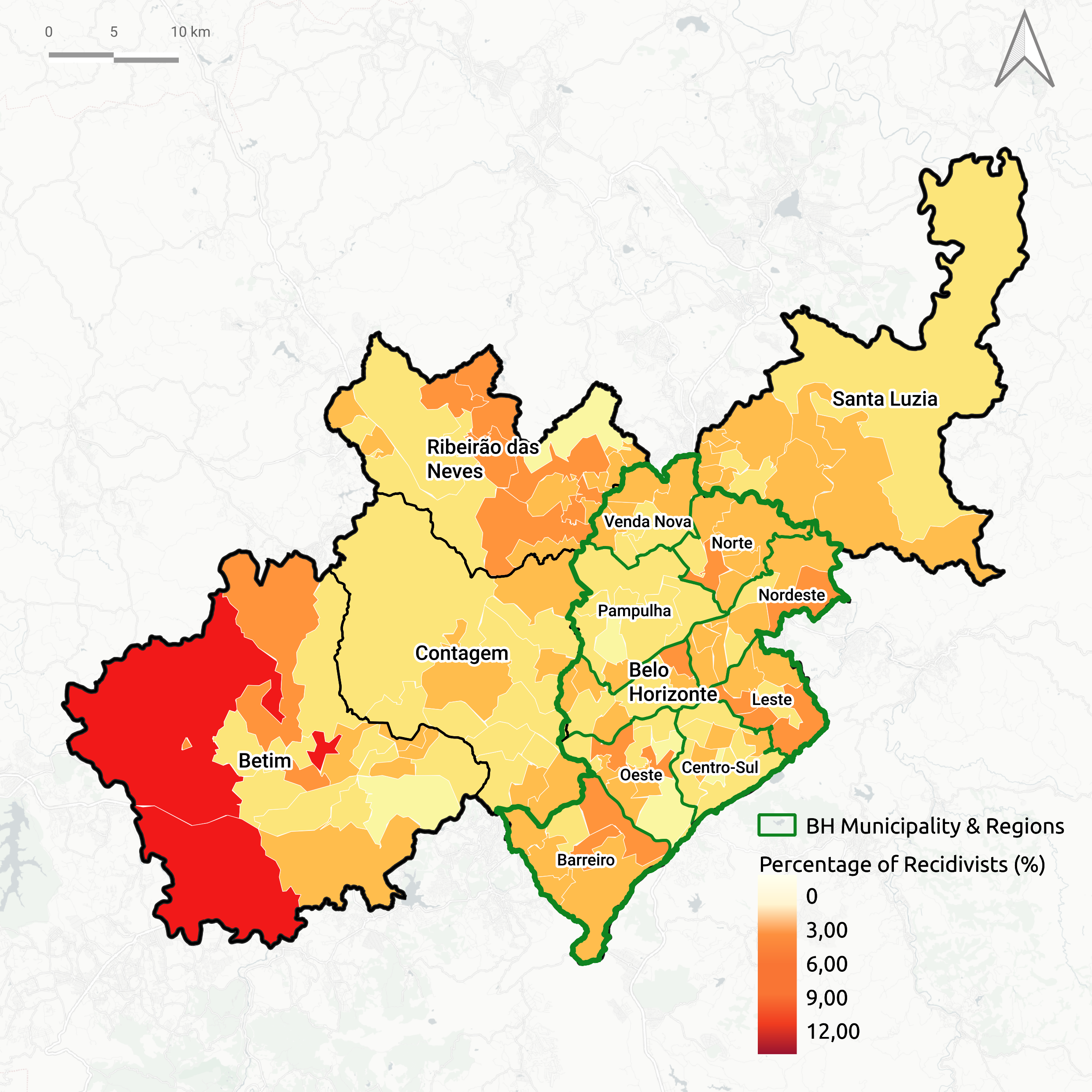}
  \caption{Percentage of individuals who reoffended among those who committed at least one crime of bodily injury;}
  \label{fig:mapa_percent_reincid}
\end{minipage}
\hfill
\begin{minipage}{0.48\linewidth}
  \centering
  \includegraphics[width=\linewidth]{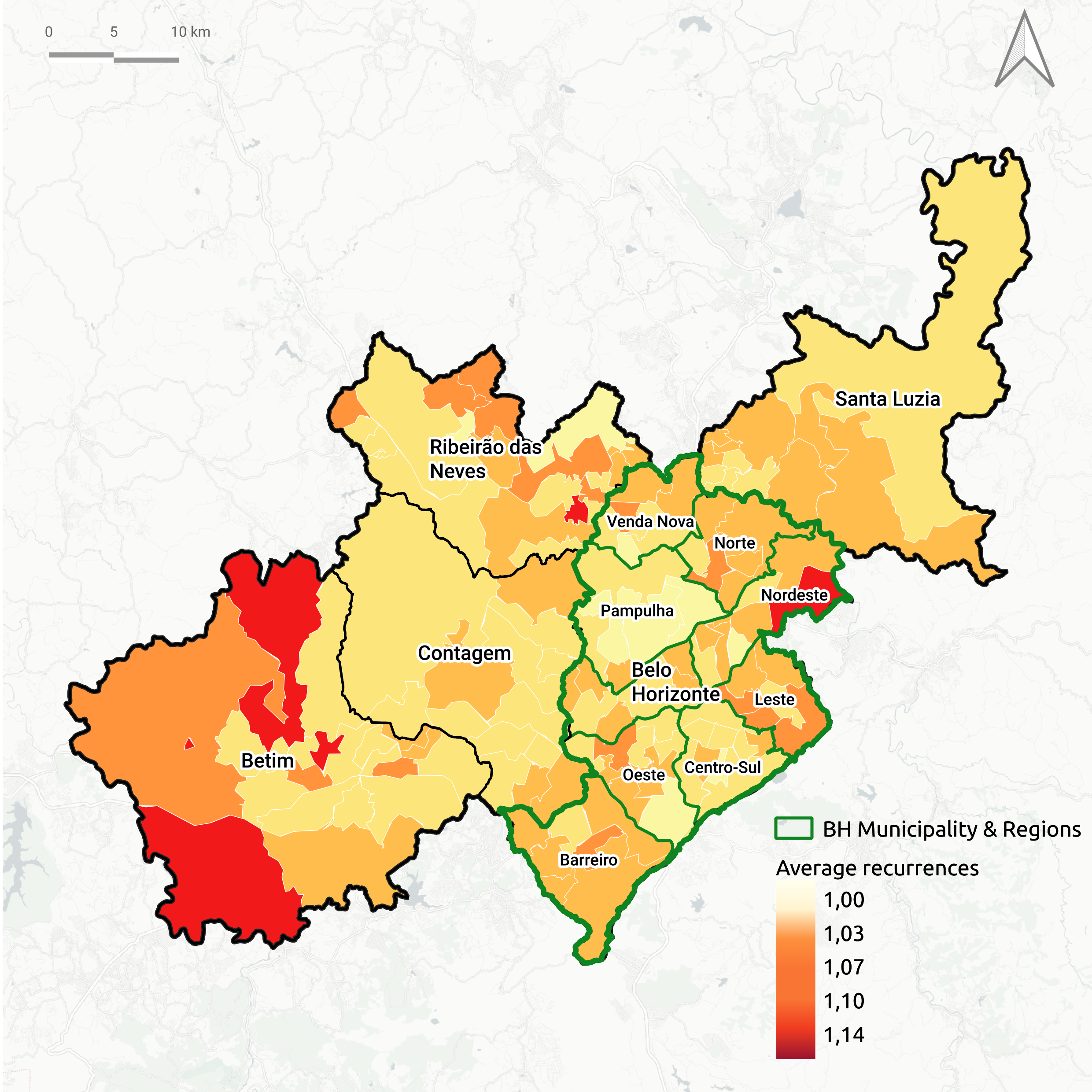}
  \caption{Average number of reoffenses among those who committed at least one crime of bodily injury}
  \label{fig:mapa_media_reincid}
\end{minipage}
\end{figure}

During the monitoring period, each individual experienced an average of 1.12 occurrences. A significant portion, corresponding to $93.31\%$ of the total, did not exhibit recidivism in relation to aggravated assault crimes, indicating the zero-inflation phenomenon within the dataset. The percentage of recidivism offenders among those who committed at least one aggravated assault crime and the average number of recidivism cases committed by these individuals, by weighting area (IBGE)\footnote{The weighting areas delimited by the Brazilian Institute of Geography and Statistics (IBGE) seek to produce relevant data for urban planning and public policies, providing an adequate structure for defining strategies aimed at public security.}, are statistics that contribute to the understanding of spatial behavior associated with recurrences of aggravated assault crimes.
Figures \ref{fig:mapa_percent_reincid} and \ref{fig:mapa_media_reincid} depict the outcomes of these statistics across selected municipalities, segmented into distinct weighting areas. An analysis of these statistics in Belo Horizonte reveals a discernible pattern: areas closer to the central region and Pampulha exhibit lower average recidivism rates and a reduced percentage of recidivism offenders. Conversely, as one moves away from these central areas, recidivism rates tend to rise, particularly in peripheral regions like the East, Northeast, North, and Barreiro areas.

The analysis, encompassing the entire region, indicates a relatively stable situation between the central region of Belo Horizonte and a section of Betim municipality. However, the westernmost area of Betim tends to exhibit higher rates. A noticeable pattern of escalation in these indicators is also evident from the central region of Belo Horizonte extending to the southern region of Santa Luzia municipality. For the most extensive weighting area in the northern region of this municipality, the indicators are among the lowest calculated for the municipality. These findings highlight the influence of spatial factors on the two indicators,  
emphasizing the necessity for the models developed in this study, as they possess structures capable of capturing the spatial dependencies inherent in the dataset while effectively addressing zero inflation concerns.

\section{Literature Review}\label{subsec2}

\cite{cook2007statistical} assert that there are two fundamental approaches to describe and model recurrent events: one involves analyzing the waiting time between successive events (referred to as ``gap time''), while the other entails event count analysis. Gap time-based analyses are more suitable when there is a renewal process for the individual after each event occurrence \citep{prentice1981regression}. On the other hand, models and methods based on event counting are more appropriate when individuals frequently experience the event of interest, and these events are incidental, meaning their occurrence does not impact the underlying process \citep{lawless2002statistical}. One common framework used for event count analysis is the Poisson process, which typically employs calendar time or process age as the time scale.

The criminal career approach acknowledges that individuals initiate their criminal activity at a specific context, become involved in crime, commit various offenses, and eventually discontinue their criminal behavior \citep{piquero2003criminal}. Consequently, an evolution in the criminal practices of recidivism offenders is expected, without the presence of a renewal process in their conduct. This aspect makes models based on event counts a suitable option for analyzing criminal recidivism data.

\subsection{The Poisson Process}

One of the structures used for event count analysis is the Non-Homogeneous Poisson Process (NHPP), which uses the intensity function, denoted by $\lambda(t)$, to provide the instantaneous marginal probability of an event occurring at time t. In the absence of covariates, the intensity function is determined exclusively by time $t$, and is represented by:
\begin{align} \label{"intensidade_geral"}
\lambda(t)=\lim_{\Delta t\to\ 0} \frac{Pr\{\Delta N(t)=1\}}{\Delta t}=\lambda(t),
\end{align}
where \(\Delta N(t)\) denotes the number of events in \([t, t+\Delta t)\).

The accumulated expected number of events up to time $t$, denoted by $\Lambda(t)$, is represented by:
\begin{align*} 
\Lambda(t)=E\{N(t)\}=\int_{0}^{t}\lambda(s)ds \ \ \ \text{for } t>0,
\end{align*}
where \(N(t)\) and the number of events occurring in the interval \((0,t]\). The Poisson process can be modeled parametrically or non-parametrically. One of the most frequently cited parametric model in the literature is the Power Law, also known as the Power Law Process \citep{rigdon2000statistical, cook2007statistical}. This model describes \(\lambda(t)\) raising time to a power and can be represented by the form:
\begin{align}\label{"PLP"}
\lambda(t;\alpha_1,\alpha_2)&=\alpha_1 \alpha_2 t^{\alpha_2-1}\ \ \ \text{$t\geq0;\alpha_1>0,\alpha_2>0$}
\end{align}
where \(\alpha_1\) is a scale parameter and \(\alpha_2\) is a shape parameter. When \(\alpha_2 = 1\), the intensity function is constant, and we have a Homogeneous Poisson Process (HPP). In the case where \(\alpha_2>1\), the intensity function is increasing and \(\alpha_2<1\) implies a decreasing intensity function. By definition \(\lambda(t)=0\) when $t=0$.

When incorporating covariates, the intensity function can take the following form:
\begin{align}  \label{"intensidade_01"}
\lambda_i(t|\boldsymbol\theta)=\lambda_0(t|\boldsymbol\alpha)\exp(\textbf{x}'_i\boldsymbol\beta)
\end{align}
where $\lambda_0(t|\boldsymbol\alpha)$ is the baseline intensity function, \(\textbf x\) is a vector of covariates, \(\boldsymbol\beta\) is a vector of parameters associated with \(\textbf x\) and \(\boldsymbol\theta=\{\boldsymbol{\alpha},\boldsymbol\beta\}\).

Methods based on the Poisson distribution assume that the mean and variance of the number of event occurrences are equal. But there are cases where this assumption does not apply. For example, datasets with a large number of zeros in the count variable, or datasets with unobserved heterogeneity, may result in a variation in the number of event occurrences that exceeds the ability of the Poisson process to model appropriately the recurrent events.

\subsection{The Poisson Process for Data with Excess of Zeros}
Studies such as those by \cite{loeber1998serious} and \cite{piquero2012violence} suggest that a notable proportion of individuals involved in criminal incidents do not exhibit subsequent recidivism, resulting in an excess of zeros regarding the number of recidivism events. Datasets with these characteristics challenge the assumptions inherent in the Poisson model, as its mean-variance relationship would not be applicable to these specific scenarios.

\cite{cook2007statistical} present the Zero-Inflated Non-homogeneous Poisson Process (ZI-NHPP) model, an approach to analyzing count data with excess of zeros based on a mixture model obtained by extending the NHPP model. Models developed for such data contain a mixture of a point mass at zero and an untruncated count distribution, similar to the case exemplified by the Zero-Inflated Poisson model \citep[ZIP,][]{lambert1992zero}.

Let us suppose \(\{N_i(t), 0 \leq t\}\) is a counting process for subject \(i\), and \(V_i\) is a latent (unobservable) random variable. When \(V_i = 1\), subject \(i\) is in a condition that allows events to occur, with probability \(Pr(V_i=1)=\pi_i\). Conversely, when \(V_i = 0\), subject \(i\) is in a condition where no events occur, resulting in a counting process that always yields zero. If \(\{N_i(t), 0 \leq t\} \mid V_i=1\) is a Poisson process with an intensity function represented by Equation \eqref{"intensidade_01"}, and \(Pr(N_i(\infty)=0 \mid V_i=0)=1\), then the marginal distribution is a mixed Poisson process that accommodates excess zeros. In this model, \(E\{N_i(t)\} = \Lambda(t)_i\pi_i\) and \(var\{N_i(t)\} = \Lambda_i(t)\pi_i + \Lambda_i^2(t)\pi_i(1-\pi_i)\), where \(\Lambda_i(t) = \int_0^t \lambda_i(s) ds\) \citep{cook2007statistical}.

Generally, a model for \(\pi_i\) is adopted, where \(g(\pi_i)=\textbf{z}'_i\boldsymbol\psi\) is a common link function for binary data. For example, \(\pi_i\) may depend on a vector of covariates \(\textbf{z}_i'\) following the logistical form presented below:
\begin{align}\label{"logistc"}
\pi_i(\textbf{z}_i')=\frac{\exp(\textbf{z}_i'\boldsymbol{\psi})}{1+\exp(\textbf{z}_i'\boldsymbol{\psi})}
\end{align}
where \(\boldsymbol{\psi}\) represents the vector of parameters associated with the vector of covariates \(\textbf{z}_i'\). Alternatively, it is possible to consider \(\pi_i\) as a constant, representing the proportion of recidivism offenders.

To define the likelihood function for the ZI-NHPP model, let us suppose that a set of \(m\) individuals remains under observation until time \(y_i\), where \(i=1, 2, \dots, m\), whether this is a failure or truncation time. Let us further suppose that, for each individual \(i\), the event of interest occurred \(n_i\) times, respectively at times \(0 \leq t_{i1} < t_{i2} <\dots < t_{in} \leq y_i\). Consider the intensity function represented in \eqref{"intensidade_01"}, denote by \(\boldsymbol\Theta=\{\boldsymbol{\theta}, \boldsymbol \psi\}\) the set of parameters to be estimated and \(D=\{(y_i,\textbf t_{ij},\textbf{x}_i,\textbf{z}_i,{n}_i): i=1,\dots,m)\}\) the observed dataset for \(m\) individuals.

To estimate the intensity function, the likelihood function for an individual \(i\) is considered based on the conditional probability density of the observed result ``\(n\) events at times \(0 \leq t_1 < t_2 <\dots < t_n \leq y\)'' which is represented by:
\begin{align} \label{"verossim_ZIP"}
L_i^1(D|\boldsymbol\Theta)\propto&\left[\prod_{j=1}^{n_i}\lambda_i(t_{ij}|\boldsymbol\theta)\exp\left\{-\int_0^{y_i}\lambda_i(s|\boldsymbol\theta)ds \right\}\pi_i\right]^{I(n_i>0)}\\
                        \times&\left[\exp\left\{-\int_0^{y_i}\lambda_i(s|\boldsymbol\theta)ds \right\}\pi_i+(1-\pi_i)\right]^{I(n_i=0)} \notag.
\end{align}

\subsection{Models with Random Effects}

An important aspect in studies on criminal careers is understanding why some individuals are more prone to crime. This propensity may depend on unobservable factors, which could be related to personal or social characteristics or a combination of both \citep{kazemian2009can}. As discussed by \cite{lawless1987regression}, the presence of this unobservable heterogeneity among individuals can lead to a variation in the number of event occurrences that is significantly greater than what is supported by the Poisson process alone. In such scenarios, it becomes essential to incorporate an individual random effect to account for this unobservable heterogeneity.

Random effects models emerge as an approach capable of modeling inter- and intra-individual variation. \citet{cook2007statistical} present the Poisson process with Random Effects, where the intensity function is specific to the individual and conditioned on its random effect, having its form represented by:
\begin{align} \label{"modelo rnd1"}
\lambda_i(t|\boldsymbol\theta,u_i) = u_i\lambda_0 (t|\boldsymbol\alpha)\exp\{\mathbf{x}'_i\boldsymbol\beta\},
\end{align}
where \(u_1,\dots,u_m\) are	 the values of the random, non-negative effects, assumed as a sample of \(U{\sim} G(u;\phi)\), where \(E(u)=1\) and \(Var(u)=\phi\). The model presented in \eqref{"modelo rnd1"} can alternatively be expressed as:
\begin{align} \label{"modelo rnd2"}
\lambda_{i}(t|\boldsymbol\theta,\omega_i) = \lambda_0 (t|\boldsymbol\alpha)\exp\{\mathbf{x}'_i\boldsymbol\beta+\omega_i\}, 
\end{align}
where \(u_i\), corresponds to \(\exp\{\omega_i\}\). In this model, it is assumed that \(\omega_i\)'s are an independent sample from some distribution with mean 0 and variance \(\sigma^2\), such as \(\omega_i{\sim} N(0,\sigma^2)\). In the case where \(\sigma^2 =0\), model \eqref{"modelo rnd2"} reduces to model \eqref{"intensidade_01"}.

Random effects are often used to represent grouping structures that may constitute a source of variation in the occurrence of events. For example, these effects may correspond to groups that are spatially organized, such as geographic or clinical regions. In the next section, the models proposed in this paper will be presented, highlighting their ability to model the spatial dependence found in the datasets.

\section{Methodology}\label{subsec2}
In this section, new models for analyzing zero-inflated and spatially correlated recurrent events data are introduced. These models, utilizing counting methods, are designed to integrate a framework of spatial random effects and to address zero-inflation in the data. Alongside the parametric versions, we will also present semiparametric models that approximate the baseline function using Bernstein polynomials. 

\subsection{The ZI-NHPP-SE and ZI-NHPP-SE-COV models}

In many studied phenomena, specific location characteristics can influence nearby areas, leading to spatial correlation, which can significantly contribute to the variation in event occurrences. Studies such as those by \citet{harries1974geography} and \citet{vandeviver2017geography} underscore the importance of incorporating the spatial dimension in the analysis of data related to crime. \citet{banerjee2003frailty}, \citet{cooner2006modelling}, \citet{bao2019semi} and \citet{adekpedjou2021semiparametric} present proposals for modeling time to event data with spatial correlation through the introduction of a spatial random effects structure. However, these studies either only considered the time until the first event occurred, or did not consider the situation in which a large portion of the population does not experience the recurrence of the event of interest.

The models proposed in this section were developed based on extending the ZI-NHPP model by incorporating a random effects structure used to model the association between individuals belonging to the same spatial stratum. To achieve this, we define $\boldsymbol{\omega}=(\omega_1,\dots,\omega_L)$ as the vector of unobservable random effects, where $\omega_l$ represents the random effect associated with the $l$-th spatial stratum, encompassing a total of $L$ individuals. Let $\boldsymbol\theta=\{\boldsymbol{\alpha},\boldsymbol\beta\}$ and $\boldsymbol\Theta=\{\boldsymbol{\theta},\boldsymbol \psi,\tau,\boldsymbol \omega\}$ denote the set of parameters to be estimated.

The formulation of the Zero-Inflated Non-Homogeneous Poisson Process with Spatial Effects (ZI-NHPP-SE) model follows from the introduction of the spatial random effect in the intensity function represented in \eqref{"modelo rnd1"}. Thus, the intensity function for an individual $i$ belonging to the $l$-th spatial stratum is defined by:
\begin{align} \label{"func_intensidade_espacial"}
\lambda_{il}(t|\boldsymbol\theta,\omega_{l})=\lambda_0 (t|\boldsymbol\alpha)\exp\{\textbf{x}'_{il}\boldsymbol\beta+\omega_{l}\},
\end{align}
where $\textbf{x}_{il}$ is a vector of covariates of individual $i$ of spatial stratum $l$.

In the context where the goal is to model the correlation among individuals within the same spatial stratum, methods for analyzing area data appear to be more appropriate. The Intrinsic Conditional Auto-Regressive Model \citep[ICAR,][]{besag1991bayesian} can be employed to introduce a spatial dependence structure among the components of the vector $\boldsymbol{\omega}$. This model is formulated based on a proximity matrix $A$ of dimension $L \times L$, where entries are set to 1 if areas $l$ and $r$ share a common border; otherwise, they are set to 0. According to \citet{banerjee2003frailty}, the complete prior conditional distribution associated with each spatial stratum can be represented as:
\begin{align*}
\omega_l|\boldsymbol{\omega}_{(-l)} \sim N \left(\frac{\sum_{l\neq j} \omega_l}{a_{l+}}, \frac{1}{\tau a_{l+}} \right), \quad l=1,\dots,L;
\end{align*}
where $\boldsymbol{\omega}{(-l)}=\left(\omega_r:r\neq l\right)$, $a_{l+}=\sum^{L}_{r=1}a_{l}$, considering that $a_{l+}$ corresponds to the number of neighbors of area $l$, and $\tau$ is a precision parameter that controls the common variability associated with the set of all areas under investigation. 
Despite being the most traditional spatial model to accommodate areal dependence, \citet{morris2019bayesian} showed that using the ICAR model significantly reduces the computational cost involved in calculating the model's probability density, leading to substantial time savings when fitting models using MCMC on datasets representing geographic regions with numerous area units.

The likelihood function for the ZI-NHPP-SE model is derived by reformulating Equation \eqref{"verossim_ZIP"}, incorporating the intensity function defined in \eqref{"func_intensidade_espacial"}, where the ICAR model establishes the dependence structure among the components of $\boldsymbol{\omega}$. Thus, \(L^1_{il}(D|\boldsymbol\Theta)\) represents the contribution of an individual \(i\), from spatial stratum \(l\), to the likelihood, where the time of the \(j\)-th occurrence of the individual \(i\), belonging to the spatial stratum \(l\) is denoted by \(t_{ijl}\). The likelihood can be expressed as:
\begin{align} \label{"verossim_ZIP3"}
L^1(D|\boldsymbol{\Theta})\propto\prod_{l=1}^{L}\prod_{i=1}^{m_l} L^1_{il}(D|\boldsymbol{\Theta}). 
\end{align}

The inclusion of the regression structure for $\pi_i$, as represented in the form described in \eqref{"logistc"}, within $L^1_{il}(D|\boldsymbol\Theta)$, defines the formulation of the Zero-Inflated Non-Homogeneous Poisson Process with Spatial Effects, with Covariates in Logistic Regression (ZI-NHPP-SE-COV) model .

\subsection{The SZI-NHPP-SE and SZI-NHPP-SE-COV Models}
An essential aspect to consider when using modeling through the Poisson process is the choice of the shape of the baseline intensity function. Semiparametric models emerge as a more flexible and robust alternative compared to purely parametric models, since they do not impose a specific form for the baseline intensity function \citep{carnicer1993shape, DEMARQUI2012728, farouki2012bernstein, osman2012nonparametric}. This characteristic allows an efficient approach to intensity functions with different shapes, expanding the applicability of these models in scenarios where the baseline intensity functions take on more complex forms than those considered by traditional parametric approaches.

The models presented in the previous section use parametric forms for the baseline intensity function, such as the Power Law, which is restricted to monotone forms. The purpose of this section is to introduce the semiparametric counterparts of the ZI-NHPP-SE and ZI-NHPP-SE-COV models, where the baseline intensity functions are approximated through the use of Bernstein Polynomials \citep{bernstein1912demo}. These new models are coined SZI-NHPP-SE and SZI-NHPP-SE-COV, respectively.

As highlighted by \citet{osman2012nonparametric}, Bernstein polynomials offer the best approximation among all polynomial alternatives, efficiently preserving the form of the desired function. \cite{bernstein1912demo} demonstrates that the approximation of a continuous function \(C(y)\) defined in the compact interval \([0,\tau]\), where \(y\) is an element of this interval, can be performed via Bernstein Polynomials of degree \(d\), represented by:
\begin{align}
B_d(y;C)=\sum_{k=0}^db_{k,d}B_{k,d}(y).
\end{align}
where the terms $B_{k,d}$ and $b_{k,d}$ are named base and coefficients\footnote{For more details, see \cite{bernstein1912demo} and \cite{lorentz1986bernstein}.} of the Bernstein polynomial of degree $d$.

\cite{Rumenickda2019modelos} employs a semiparametric approach, where the baseline intensity function is obtained using Bernstein polynomials. Following the framework outlined by \citet{osman2012nonparametric}, the authors demonstrate that the intensity function can be approximated as:
\begin{align*}
\lambda_d(t|\Lambda)&=\sum_{k=1}^d\left[\Lambda\left(\frac{k}{d}\zeta \right)-\Lambda\left(\frac{(k-1)}{d}\zeta\right)\right]\times \frac{f_{Beta}\left(t/\zeta|k,d-k+1\right)}{\zeta},
\end{align*}
where $0<T_1<T_2<\dots<\zeta$ are random variables that represent the times of occurrence of events, $\Lambda(.)$ is the accumulated intensity function and $f_{Beta}$ is the probability density function of the Beta distribution with parameters $k$ and $d-k+1$ , evaluated at the point $t/\zeta$.

By definition, $\gamma_k = \Lambda\left(\frac{k}{d}\zeta\right) - \Lambda\left(\frac{(k-1)}{d}\zeta\right)$ and $g_{d,k} = \frac{f_{Beta}(t/\zeta|k, d-k+1)}{\zeta}$. The intensity function, as expressed in Equation \eqref{"intensidade_geral"}, is induced using the Bernstein Polynomials approach as follows:
\begin{align}\label{"intensidade_PB_rumenick_0"}
\lambda_d(t|\gamma)=\sum_{k=1}^d\gamma_k g_{d,k}(t)=\boldsymbol{\gamma}' \boldsymbol{g}_d(t),
\end{align}
where $\boldsymbol{\gamma}=(\gamma_1,\gamma_2,\dots,\gamma_d)'$ represents a vector of unknown coefficients with $\gamma_k \geq 0$, and $\boldsymbol{g}_d(t)=(g_{d,1}(t),\dots,g_{d,d}(t))'$ denotes a vector of basis functions. Additionally, let $\boldsymbol\beta$ be a vector of regression coefficients associated with the intensity function and $\boldsymbol\Theta=\{\boldsymbol\theta\}$ be the set of parameters to be estimated, where $\boldsymbol\theta=\{\boldsymbol{\gamma},\boldsymbol\beta\}$ represents the parameters related to the intensity function. Moreover, $D=\{(y_i,\textbf t_{ij},\textbf{x}_i,\textbf{z}_i,{n}_i): i=1,\dots,m)\}$ stands for the observed dataset containing data for \(m\) individuals. The intensity function of the Poisson process using Bernstein polynomials is expressed as:
\begin{align}  \label{"intensidade_PB_rumenick"}
\lambda_i(t|\boldsymbol\theta)=\lambda_d(t|\boldsymbol\gamma)\exp(\textbf{x}'_i\boldsymbol\beta),
\end{align}
where \(\lambda_d(t_{ij}|\boldsymbol{\gamma})\) is the base intensity function induced by PBs defined in \eqref{"intensidade_PB_rumenick_0"}. The expected accumulated number of events, $\Lambda(t)$, is obtained by:
\begin{align*} 
\Lambda_d(y_i|\boldsymbol\gamma) = \int_0^{y_i}\lambda_d (s|\boldsymbol\gamma)ds = \boldsymbol\gamma'\boldsymbol{G}_d(y_i), 
\end{align*}
where $\boldsymbol{G}_d(t)=(G_{d,1}(t),G_{d,2}(t),\dots,G_{d,d}(t))'$, with $G_{d,k}(t)=\int_{0}^{t}g_{d,k}(u)du=F_{Beta}(t/\zeta|k, d-k+1)$, for $k=1,2,\ldots,d$. $F_{Beta}$ is the cumulative distribution function of Beta with parameters $k$ and $d-k+1$, evaluated at point $t/\zeta$.

The formulation of the Semiparametric Zero-Inflated Non-Homogeneous Poisson Process with Spatial Effects (SZI-NHPP-SE) model involves incorporating a baseline intensity function induced by PBs into the model specified in Equation \eqref{"func_intensidade_espacial"}. To achieve this, let $\boldsymbol\theta=\{\boldsymbol{\gamma},\boldsymbol\beta\}$ and $\boldsymbol\Theta=\{\boldsymbol{\theta},\boldsymbol \psi,\zeta,\boldsymbol{\omega}\}$ represent the set of parameters to be estimated, where $\boldsymbol{\gamma}=(\gamma_1, \dots, \gamma_d)'$ are the unknown coefficients of polynomials of degree $d$. Additionally, consider that the intensity function is represented as follows:
\begin{align} \label{"FI_BP_1"}
\lambda_{il}(t|\boldsymbol \theta,\omega_l)=\lambda_{d} (t|\boldsymbol\gamma)\exp\{\textbf{x}'_{il}\boldsymbol\beta+\omega_{l}\},
\end{align}
where \(\lambda_{d}(t|\boldsymbol{\gamma})\) is the base intensity function induced by the PBs defined in Equation \eqref{"intensidade_PB_rumenick_0"}, and \(\omega_l\), as specified in Equation \eqref{"func_intensidade_espacial"}, represents the random effect associated with the \(l\)-th spatial stratum. Similar to the definition in Equation \eqref{"verossim_ZIP3"}, the contribution of an individual \(i\), belonging to spatial stratum \(l\), to the likelihood function is denoted by \(L^1_{il}(D|\boldsymbol\Theta)\), but with an intensity function represented by Equation \eqref{"FI_BP_1"}. The likelihood can be expressed as:
\begin{align*} 
L_d^1(D|\boldsymbol{\Theta})\propto\prod_{l=1}^{r}\prod_{i=1}^{m_l} L^1_{il}(D|\boldsymbol{\Theta}). 
\end{align*}

In this approach, the random effects presented in Equation \eqref{"FI_BP_1"} reflect variations relative to area units. These effects are subject to a correlation structure represented by the ICAR model. 
The formulation of the Semiparametric Zero-Inflated Non-Homogeneous Poisson Process with Spatial Effects, with Covariates in Logistic Regression (SZI-NHPP-SE-COV) model is due to the inclusion of the regression structure for \(\pi_i\) represented in the form described in \eqref{"logistc"}.

\section{Inference}\label{subsec2}

Inference for the models presented in this paper will be conducted from a Bayesian perspective, which provides the advantage of its ability to formally incorporate external empirical evidence into the results through the prior distribution. Furthermore, as highlighted by \citet{banerjee2003hierarchical}, in the context where \(\boldsymbol{\omega}\) is intended to be modeled as random effects, the Bayesian approach enables the establishment of specific correlation structures among these effects and between the observed data \(y_i\).

In the Bayesian framework, inferences regarding quantities of interest rely on their posterior joint distributions. However, for the models introduced in this paper, obtaining these distributions analytically is not feasible. Therefore, to acquire samples from these posterior joint distributions, Markov Chain Monte Carlo (MCMC) methods such as the Hamiltonian Monte Carlo (HMC) algorithm are employed \citep{alder1959studies, migon2014statistical}. \texttt{Stan} is a statistical modeling software that employs the HMC method to generate representative samples from posterior distributions. In this study, \texttt{Stan} was utilized via the \texttt{rstan} package \citep{rstan}, enabling the utilization of the \texttt{Stan} language within an \texttt{R} environment \citep{R}.

Model specification in the Bayesian context requires eliciting prior distributions for the model parameters. Assuming that the baseline intensity function follows the Power Law form, as presented in Equation \eqref{"PLP"}, a common choice for both the shape parameter and the scale parameter is the gamma distribution \citet{rigdon2000statistical}. This distribution encompasses a range of shapes and is supported by positive real numbers.

For the intensity function induced by Bernstein Polynomials, the parameters in $\boldsymbol\gamma$ have positive support, however, in this study, they are modeled on the logarithmic scale due to computational reasons. Thus, we have $\log(\gamma_k) \sim N(\mu_\gamma, \sigma_\gamma)$. However, other prior distributions can be used, which may induce some type of dependency among their components.

For representing subjective information associated with regression coefficients, the normal distribution is commonly employed. Therefore, the following prior distributions are considered for the parameters of the models proposed in this work: $\alpha_1\sim Gamma(a_{\alpha_1},b_{\alpha_1})$; $\alpha_2\sim Gamma(a_{\alpha_2},b_{\alpha_2})$; $\beta_l\sim N(\mu_{\beta},\sigma^2_{\beta})$, com $l=0,\ldots,p$; $\psi_m\sim N(\mu_{\psi},\sigma^2_{\psi})$, com $m=0,\ldots,q$; $\log(\gamma_k) \sim N(\mu_\gamma, \sigma_\gamma)$, com $k=1,\ldots,d$; $\tau\sim Gamma(a_{\tau},b_{\tau})$; $\pi\sim Beta(a_{\tau},b_{\tau})$. The posterior joint distributions of the proposed models are expressed as:
\begin{itemize}
    \item ZI-NHPP-SE: $p(\boldsymbol{\Theta}|D)\propto 
L^1(D|\boldsymbol{\Theta})p(\boldsymbol\omega|\boldsymbol\tau)p(\boldsymbol\alpha)p(\boldsymbol\beta)p(\boldsymbol\pi)p(\boldsymbol\tau)$,;
    \item ZI-NHPP-SE-COV: $p(\boldsymbol{\Theta}|D)\propto 
L^1(D|\boldsymbol{\Theta})p(\boldsymbol\omega|\boldsymbol\tau)p(\boldsymbol\alpha)p(\boldsymbol\beta)p(\boldsymbol\psi)p(\boldsymbol\tau)$;
     \item SZI-NHPP-SE-COV: $p(\boldsymbol{\Theta}| \boldsymbol\omega , D)\propto 
L_d^1(D|\boldsymbol{\Theta})p(\boldsymbol\omega|\boldsymbol\tau)p(\boldsymbol\gamma)p(\boldsymbol\beta)p(\boldsymbol\pi)p(\boldsymbol\tau)$;
      \item SZI-NHPP-SE-COV: $p(\boldsymbol{\Theta}| \boldsymbol\omega , D)\propto 
L_d^1(D|\boldsymbol{\Theta})p(\boldsymbol\omega|\boldsymbol\tau)p(\boldsymbol\gamma)p(\boldsymbol\beta)p(\boldsymbol\psi)p(\boldsymbol\tau)$;
\end{itemize}
where $p(.)$ represents the prior distribution of a given parameter.

\subsection{Model Selection}\label{subsec2}
Bayesian model comparison will be performed using the Widely Applicable Information Criterion (WAIC) and the Pareto Smoothed Importance Sampling - Leave-One-Out Cross-Validation (PSIS-LOO) methods.

\subsubsection{\texorpdfstring{WAIC}{WAIC}}\label{waic-widely-applicable-information-criterion}

The WAIC criterion, introduced by \citet{watanabe2010asymptotic}, offers a unique approach compared to other criteria as it directly measures in-sample predictive accuracy without approximation. Mathematically, the WAIC criterion is expressed as:
\begin{align}
WAIC=\tilde{A}(x)-p{W}=\sum_{i=1}^{n}\ln E_{\theta|x}[p(x_i|\theta)]-p_W. \notag
\end{align}

One possible formulation for \(p_w\) is:
\begin{align}
p_{W}&\simeq-2\sum_{i=1}^{n}\left \{\frac{1}{m}\sum_{j=1}^{m}\ln p(x_i|\theta_{(j)})-\ln \left[\frac{1}{m}\sum_{j=1}^{m}p(x_i|\theta_{(j)})\right]\right\}, \notag
\end{align}
where \(p_{w}\geq 0\).

Another variant proposed by \citet{gelman2014understanding}, which is on the same scale as the deviance for ease of comparison, is represented as:
\begin{align}
WAIC=-2\sum_{i=1}^{n}\ln E_{\theta|x}[p(x_i|\theta)]+2p_W. \notag
\end{align}

In \citet{gelman2014understanding}'s interpretation, the term \(p_W\) serves as an approximation of the number of "free" parameters in the model. Each parameter is counted as 1 if it can be estimated without constraints or prior information and 0 if its information is solely derived from restrictions or prior knowledge. In essence, the effective model dimension is influenced by both the available data and the underlying prior distribution.

\subsubsection{\texorpdfstring{PSIS-LOO}{PSIS-LOO}}\label{Leave-One-Out-Cross-Validation}

The PSIS-LOO technique, proposed by \citet{vehtari2017practical}, estimates the Bayesian LOO predictive fit (\emph{Leave-one-out cross-validation}) through the formula: $\tilde{A}{loo} = \sum{i=1}^n \log p(y_i|y_{-i})$, where $ p(y_i|y_{-i}) $ represents the LOO predictive density considering the data without the $i$-th data point. This density can be approximated using samples $\boldsymbol{\theta}^s$ from the complete posterior distribution $ p(\boldsymbol{\theta}|y) $ with importance indices $r_i^s$ given by:
\begin{equation*}
r_i^s = \frac{1}{p(y_i|\boldsymbol{\theta}^s)} \propto \frac{p(\boldsymbol{\theta}^s|y_{-i})}{p(\boldsymbol{\theta}^s|y)}.
\end{equation*}

To obtain the leave-one-out (IS-LOO) predictive distribution, importance sampling is used:
\begin{equation*}
p(\tilde{y}_i|y_{-i}) \approx \frac{\sum_{s=1}^S r_i^s p(\tilde{y}_i| \theta^s)}{\sum_{s=1}^S r_i}.
\end{equation*}

The evaluation of the LOO predictive logarithmic density at the retained data point $y_i$  is given by:
\begin{equation*}
p(y_i|y_{-i}) \approx \frac{1}{\frac{1}{S}\sum_{s=1}^S \frac{1}{p(y_i|\theta^s)}}.
\end{equation*}

However, there's a potential issue when directly using importance indices because the posterior distribution $p(\theta|y)$ might exhibit a smaller variance and narrower tails compared to the leave-one-out distributions $p(\theta|y_{-i})$. This discrepancy can lead to instability. To address this concern, \citet{koopman2009testing} suggested an approach that involves adjusting the generalized Pareto distribution to the upper tail of the distribution of importance indices through maximum likelihood estimation. Additionally, they developed a test to assess the finiteness of the variation in importance indices.

\citet{ionides2008truncated} proposes a modification of the importance index using a truncated weight that guarantees finite variation: $w_i^s = \min(r_i^s, \sqrt{S}\bar{r})$. \citet{vehtari2015pareto} introduced Pareto Smoothed Importance Sampling (PSIS) to enhance Bayesian LOO estimation by smoothing the importance indices. PSIS-LOO is asymptotically equivalent to WAIC but exhibits greater robustness in finite cases, particularly with uninformative prior distributions or influential observations. The PSIS-LOO expression is given by:
\begin{equation*}
\widehat{PSIS-LOO} = -2\sum_{i=1}^n \log \left( \frac{\sum_{s=1}^{S}w_i^s p(y_i|\theta^s)}{\sum_{s=1}^{S}w_i^s} \right).
\end{equation*}
The estimate of the shape parameter $k$ of the generalized Pareto distribution can be used to assess the reliability of the estimate. If $k < \frac{1}{2}$, the variance of the importance indices is finite, and convergence is fast. If $\frac{1}{2} < k < 1$, the variance is infinite, but the mean exists, and convergence is slower. If $k > 1$, the variance and mean do not exist, and the variance of the PSIS estimate can be large. It is recommended to use other measures or more robust models when $k > 0.7$.

\section{Simulation Study}\label{subsec2}

This section will present the results of the simulation study conducted to evaluate the proposed models, divided into three subsections. Initially, the models will be assessed based on their capacity to accurately recover the true values of the parameters used in generating the simulated data. Subsequently, a comparative analysis will be conducted, evaluating the quality of fit of the semiparametric models against the parametric models. Finally, the ability of the models to adjust baseline intensity functions in a more flexible way than the parametric one will be evaluated.

The simulated datasets were designed to reflect certain aspects observed in the dataset on criminal recidivism in the Metropolitan Region of Belo Horizonte. To define the spatial framework, we utilized the shapefile of the region described in Section 2, which includes Belo Horizonte and four neighboring municipalities, subdivided into weighting areas, as shown in Figure \ref{fig:mapa_percent_reincid}.
For the initial evaluations, different scenarios are considered, combining varying sample sizes and average numbers of expected events, denoted as $C_{n}^{E\{N(t)\}}$. We consider sample sizes of $n=300$, $n=500$, and $n=1000$. The different expected event counts are achieved by adjusting the shape parameters of the baseline intensity function, assuming the Power Law process as presented in Equation \eqref{"PLP"}. Setting $\alpha_1=0.50$ and $\alpha_2=1.30$ yields $E\{N(t)\}=6.3$. With $\alpha_1=1.00$ and $\alpha_2=1.30$, the expected event count is $E\{N(t)\}=12.5$, and for $\alpha_1=2.00$ and $\alpha_2=1.30$, $E\{N(t)\}=25.0$.

In this study, three covariates are considered: $x_{i1}$, $x_{i2}$, and $z_{i1}$. The first two are related to the predictive part of the intensity function, while the third is related to the probability of recidivism. Covariate $x_{i1}$ is generated from a Bernoulli distribution with a success probability of $0.70$, while $x_{i2}$ and $z_{i1}$ are generated from a standard normal distribution. The coefficients of the explanatory variables were set to $\beta_1=0.60$, $\beta_2=0.80$, and $\psi_1=1.00$.

To generate the spatial random effect component values, we adopt the ICAR model structure \citep{besag1991bayesian,assunccao2002bayesian}, with precision parameter $\tau = 1$. The spatial structure is defined based on the geographic division depicted in Figure \ref{fig:mapa_media_reincid}. Event occurrence times are determined using the method proposed by \citet{jahn2015simulating}. Their work outlines an approach for simulating recurrent events data, employing the \citet{andersen1982cox} model with a total time scale. Since the dataset on criminal recidivism lacks details regarding terminal events or censures during the follow-up period, we assume uninformative right censoring for all individuals at the end of the follow-up period in this simulation study.

The results showcased in this paper were produced using the \texttt{spnhppzi} package, which can be downloaded from \href{https://github.com/alissonccs/spnhppzi}{https://github.com/alissonccs/spnhppzi}. The package was developed within the \texttt{R} software environment, leveraging the capabilities of the \texttt{rstan} package. For all datasets, $2$ chains were fitted, each with $2,000$ iterations and a burn-in of $1,000$, resulting in a final posterior chain of size $2,000$. The convergence of the chains was checked using the \emph{R-Hat} diagnostic tool \citep{vehtari2021rank}.

\subsection{Evaluation of Parametric Models}\label{subsec2}
This section aims to assess the parametric models' capability to accurately estimate the true parameter values. The analysis assumes identical structures for both the data generating models and the fitting models. We perform 300 Monte Carlo replications for each scenario to evaluate the properties of the parameter estimators in the proposed models. For a generic parameter $\varphi$ and its posterior estimate $\hat{\varphi}$, with $R$ denoting the total number of Monte Carlo replications, several statistics are utilized to evaluate the results: the average estimate (Average est.), expressed by \(\bar{\varphi}=\frac{1}{R}\sum_{i=1}^{R}\hat{\varphi}_i\); the mean of the estimated standard error (SE), represented by \(\bar{se}=\frac{1}{R}\sum_{i=1}^{R}\hat{se}(\hat{\varphi}_i)\); the standard error of estimates (SDE), calculated using \(sde=\frac{1}{R-1}\sum_{i=1}^{R}(\hat{\varphi}_i-\bar{\varphi})^2\); and the relative bias (RB), determined by \(rb(\varphi)=100(\hat\varphi-\varphi)/|\varphi|\).

Table \ref{tab:tab1} displays the results\footnote{For additional scenarios, refer to 
\href{https://alissonccs.shinyapps.io/DASH\_TABLE/}{https://alissonccs.shinyapps.io/DASH\_TABLE/}.} for the $n=300$ scenarios: $C_{300}^{6.3}$, $C_{300}^{12.5}$, and $C_{300}^{25}$. Overall, the findings demonstrate the models' strong capability to accurately estimate the true parameter values. Examination of the average relative bias reveals that, except for the spatial precision parameter $\tau$ in the scenario with lower data volume, $C_{300}^{3.6}$, all other values are below $11.00\%$.

The significance of increasing the number of expected events to reduce relative biases becomes apparent. Consider the parameter $\tau$ of the ZI-NHPP-SE model: as the number of expected events increases from \(E\{N(t)\}=6.30\) to \(E\{N(t)\}=12.50\), the relative bias decreases from $20.32\%$ to $11.83\%$. This trend is not exclusive to $\tau$ but extends to other parameters of both the ZI-NHPP-SE and ZI-NHPP-SE-COV models. Notably, in the scenario with the highest number of expected events, $C_{300}^{25}$, all observed bias values are under $7.90\%$.

When evaluating the other statistics, the results also indicate the superior performance of the models in scenarios with a greater number of expected events. The reduction in estimated values is evident for most parameters, both for the estimated standard error and the standard error of estimates. Concerning the coverage rate, it is apparent that as the average number of expected events increases, the rates approach the nominal value of $95\%$ for most cases.

\begin{table*}[h]
\caption{Mean, estimated standard error, standard error of estimates, average relative bias (\%) and coverage rate (\%) for the models in scenarios $\boldsymbol{C_{300}^{6.3}}$, $\boldsymbol{C_{300}^{12.5}}$ and $\boldsymbol{C_{300}^{25}}$.\label{tab:tab1}}
\tabcolsep=0pt
\begin{tabular*}{\textwidth}{@{\extracolsep{\fill}}llllcccrr@{\extracolsep{\fill}}}
\toprule%
\textbf{Scenarios} &\textbf{Model} & \textbf{Parameter} & \textbf{Prior Dist.} & \textbf{Average Est.} & \textbf{SE} & \textbf{SDE} & \textbf{RB (\%)} & \textbf{CR (\%)}\\
\midrule
 &  & $\alpha_1 = 0.50 $ & Gama(0.1 , 0.1) & 0.53 & 0.0030 & 0.1955 & 5.46 & 94.33\\
\nopagebreak
 &  & $\alpha_2 = 1.30 $ & Gama(0.1 , 0.1) & 1.30 & 0.0008 & 0.1176 & 0.16 & 95.33\\
\nopagebreak
 &  & $\beta_1 = 0.60 $ & N(0 , 4) & 0.59 & 0.0052 & 0.1579 & -1.60 & 94.33\\
\nopagebreak
 &  & $\beta_2 = 0.80 $ & N(0 , 4) & 0.78 & 0.0024 & 0.0371 & -2.46 & 94.00\\
\nopagebreak
 &  & $\pi = 0.75 $ & Beta(1 , 1) & 0.76 & 0.0005 & 0.0415 & 1.31 & 90.67\\
\nopagebreak
 & \multirow{-6}{*}{\raggedright\arraybackslash ZI-NHPP-SE} & $\tau = 1.00 $ & Gamma(0.01 , 0.01) & 1.20 & 0.0208 & 0.2145 & 20.32 & 94.33\\
\cline{2-9}\pagebreak[0]
 &  & $\alpha_1 = 0.50 $ & Gama(0.1 , 0.1) & 0.53 & 0.0021 & 0.1910 & 5.72 & 92.00\\
\nopagebreak
 &  & $\alpha_2 = 1.30 $ & Gama(0.1 , 0.1) & 1.30 & 0.0006 & 0.1151 & -0.06 & 94.83\\
\nopagebreak
 &  & $\beta_1 = 0.60 $ & N(0 , 4) & 0.58 & 0.0034 & 0.1576 & -3.64 & 92.67\\
\nopagebreak
 &  & $\beta_2 = 0.80 $ & N(0 , 4) & 0.78 & 0.0016 & 0.0338 & -1.98 & 92.67\\
\nopagebreak
 &  & $\psi_0 = 0.80 $ & N(0 , 4) & 0.88 & 0.0030 & 0.0308 & 10.37 & 91.83\\
\nopagebreak
 &  & $\psi_1 = 1.20 $ & N(0 , 4) & 1.16 & 0.0033 & 0.0666 & -3.58 & 93.33\\
\nopagebreak
\multirow{-13}{*}{\raggedright\arraybackslash \textbf{$\boldsymbol{C_{300}^{6.3}}$}} & \multirow{-7}{*}{\raggedright\arraybackslash ZI-NHPP-SE-COV} & $\tau = 1.00 $ & Gamma(0.01 , 0.01) & 1.13 & 0.0125 & 0.1192 & 13.47 & 92.67\\
\cmidrule{1-9}\pagebreak[0]
 &  & $\alpha_1 = 1.00 $ & Gama(0.1 , 0.1) & 1.01 & 0.0047 & 0.0155 & 1.35 & 93.67\\
\nopagebreak
 &  & $\alpha_2 = 1.30 $ & Gama(0.1 , 0.1) & 1.30 & 0.0005 & 0.1174 & 0.20 & 94.67\\
\nopagebreak
 &  & $\beta_1 = 0.60 $ & N(0 , 4) & 0.60 & 0.0041 & 0.1462 & -0.70 & 92.33\\
\nopagebreak
 &  & $\beta_2 = 0.80 $ & N(0 , 4) & 0.80 & 0.0019 & 0.0298 & -0.53 & 95.67\\
\nopagebreak
 &  & $\pi = 0.75 $ & Beta(1 , 1) & 0.75 & 0.0005 & 0.0451 & 0.16 & 93.00\\
\nopagebreak
 & \multirow{-6}{*}{\raggedright\arraybackslash ZI-NHPP-SE} & $\tau = 1.00 $ & Gamma(0.01 , 0.01) & 1.12 & 0.0133 & 0.1252 & 11.83 & 94.33\\
\cline{2-9}\pagebreak[0]
 &  & $\alpha_1 = 1.00 $ & Gama(0.1 , 0.1) & 1.02 & 0.0031 & 0.0099 & 1.67 & 94.24\\
\nopagebreak
 &  & $\alpha_2 = 1.30 $ & Gama(0.1 , 0.1) & 1.30 & 0.0004 & 0.1155 & 0.02 & 95.76\\
\nopagebreak
 &  & $\beta_1 = 0.60 $ & N(0 , 4) & 0.60 & 0.0025 & 0.1401 & -0.79 & 94.55\\
\nopagebreak
 &  & $\beta_2 = 0.80 $ & N(0 , 4) & 0.80 & 0.0012 & 0.0290 & -0.57 & 93.03\\
\nopagebreak
 &  & $\psi_0 = 0.80 $ & N(0 , 4) & 0.84 & 0.0028 & 0.0397 & 4.45 & 95.15\\
\nopagebreak
 &  & $\psi_1 = 1.20 $ & N(0 , 4) & 1.21 & 0.0031 & 0.0908 & 0.56 & 95.45\\
\nopagebreak
\multirow{-13}{*}{\raggedright\arraybackslash \textbf{$\boldsymbol{C_{300}^{12.5}}$}} & \multirow{-7}{*}{\raggedright\arraybackslash ZI-NHPP-SE-COV} & $\tau = 1.00 $ & Gamma(0.01 , 0.01) & 1.07 & 0.0087 & 0.0691 & 7.32 & 94.55\\
\cmidrule{1-9}\pagebreak[0]
 &  & $\alpha_1 = 2.00 $ & Gama(0.1 , 0.1) & 2.01 & 0.0085 & 1.1337 & 0.51 & 96.67\\
\nopagebreak
 &  & $\alpha_2 = 1.30 $ & Gama(0.1 , 0.1) & 1.30 & 0.0003 & 0.1150 & -0.04 & 94.67\\
\nopagebreak
 &  & $\beta_1 = 0.60 $ & N(0 , 4) & 0.60 & 0.0033 & 0.1376 & 0.11 & 93.67\\
\nopagebreak
 &  & $\beta_2 = 0.80 $ & N(0 , 4) & 0.80 & 0.0015 & 0.0272 & 0.11 & 95.00\\
\nopagebreak
 &  & $\pi = 0.75 $ & Beta(1 , 1) & 0.75 & 0.0004 & 0.0465 & -0.28 & 94.33\\
\nopagebreak
 & \multirow{-6}{*}{\raggedright\arraybackslash ZI-NHPP-SE} & $\tau = 1.00 $ & Gamma(0.01 , 0.01) & 1.08 & 0.0103 & 0.0765 & 7.86 & 93.67\\
\cline{2-9}\pagebreak[0]
 &  & $\alpha_1 = 2.00 $ & Gama(0.1 , 0.1) & 2.01 & 0.0053 & 1.1192 & 0.45 & 95.67\\
\nopagebreak
 &  & $\alpha_2 = 1.30 $ & Gama(0.1 , 0.1) & 1.30 & 0.0003 & 0.1160 & 0.09 & 94.67\\
\nopagebreak
 &  & $\beta_1 = 0.60 $ & N(0 , 4) & 0.60 & 0.0018 & 0.1348 & -0.18 & 94.00\\
\nopagebreak
 &  & $\beta_2 = 0.80 $ & N(0 , 4) & 0.80 & 0.0008 & 0.0277 & -0.31 & 95.33\\
\nopagebreak
 &  & $\psi_0 = 0.80 $ & N(0 , 4) & 0.82 & 0.0026 & 0.0457 & 2.00 & 94.33\\
\nopagebreak
 &  & $\psi_1 = 1.20 $ & N(0 , 4) & 1.22 & 0.0030 & 0.0975 & 1.79 & 94.67\\
\nopagebreak
\multirow{-13}{*}{\raggedright\arraybackslash \textbf{$\boldsymbol{C_{300}^{25}}$}} & \multirow{-7}{*}{\raggedright\arraybackslash ZI-NHPP-SE-COV} & $\tau = 1.00 $ & Gamma(0.01 , 0.01) & 1.06 & 0.0069 & 0.0578 & 5.58 & 92.67\\
\bottomrule
\end{tabular*}
\end{table*}

\begin{figure*}[h]%
\centering
\includegraphics[width=0.7\textwidth]{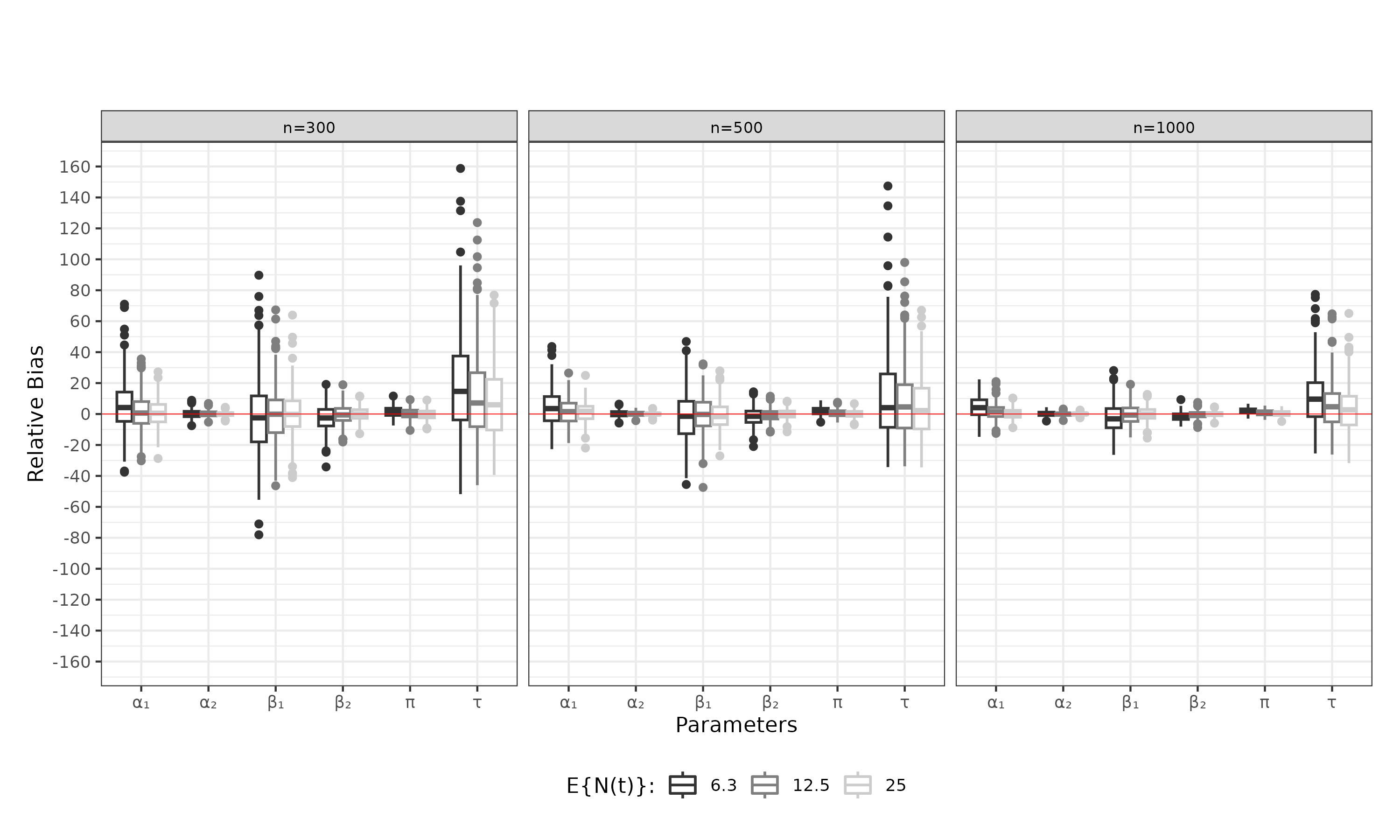}
\caption{Schematic diagrams for the estimated relative biases of the parameters in the ZI-NHPP-SE model, by sample size and expected number of events - Data generating model: PLP }\label{fig:box1}
\end{figure*}

Figure \ref{fig:box1} presents the distribution of relative biases estimated through schematic diagrams generated for the ZI-NHPP-SE model. It is observed that, for this model, the median of relative biases is less than $16\%$, and they decrease as the expected number of events increases. In scenarios with \(E\{N(t)\}=12.50\) or \(E\{N(t)\}=25\), the biases are distributed around zero for all parameters, considering the different sample sizes. Additionally, we can observe that as the sample size and the number of expected events increase, the model produces biases with less dispersion, indicating an improvement in the accuracy of the statistics generated. For the ZI-NHPP-SE-COV model, the results follow a similar pattern to the ZI-NHPP-SE model, as illustrated in the schematic diagrams available at \href{https://alissonccs.shinyapps.io/DASH_SCHEMATIC_DIAGRAMS/}{https://alissonccs.shinyapps.io/DASH\_SCHEMATIC\_DIAGRAMS/}. However, the relative bias values are smaller, and the medians are closer to zero.

The results shown in Table \ref{tab:tab2} illustrate how increasing sample size influences model estimates. Comparing scenarios \(C_{500}^{6.3}\) and \(C_{1000}^{25}\) reveals a reduction in estimated standard errors and standard errors of estimates compared to the results in Table \ref{tab:tab1}. All values associated with each parameter of all models decrease as the sample size increases from \(n=300\) to \(n=500\). Furthermore, there is a decrease in the absolute values of relative biases with increasing sample size, although this effect is less significant than the impact of the expected number of events.

\begin{table*}[h]
\caption{Mean, estimated standard error, standard error of estimates, average relative bias (\%) and coverage rate (\%) for the models in scenarios $\boldsymbol{C_{500}^{6.3}}$ and $\boldsymbol{C_{1000}^{25}}$.\label{tab:tab2}}
\tabcolsep=0pt
\begin{tabular*}{\textwidth}{@{\extracolsep{\fill}}llllcccrr@{\extracolsep{\fill}}}
\toprule%
\textbf{Scenarios} &\textbf{Model} & \textbf{Parameter} & \textbf{Prior Dist.} & \textbf{Average Est.} & \textbf{SE} & \textbf{SDE} & \textbf{RB (\%)} & \textbf{CR (\%)}\\
\midrule
 &  & $\alpha_1 = 0.50 $ & Gama(0.1 , 0.1) & 0.52 & 0.0019 & 0.1987 & 3.98 & 94.00\\
\nopagebreak
 &  & $\alpha_2 = 1.30 $ & Gama(0.1 , 0.1) & 1.30 & 0.0005 & 0.1173 & 0.17 & 95.00\\
\nopagebreak
 &  & $\beta_1 = 0.60 $ & N(0 , 4) & 0.59 & 0.0030 & 0.1503 & -2.37 & 95.00\\
\nopagebreak
 &  & $\beta_2 = 0.80 $ & N(0 , 4) & 0.79 & 0.0014 & 0.0325 & -1.62 & 92.00\\
\nopagebreak
 &  & $\pi = 0.75 $ & Beta(1 , 1) & 0.76 & 0.0003 & 0.0395 & 1.85 & 89.67\\
\nopagebreak
 & \multirow{-6}{*}{\raggedright\arraybackslash ZI-NHPP-SE} & $\tau = 1.00 $ & Gamma(0.01 , 0.01) & 1.11 & 0.0112 & 0.0963 & 10.90 & 94.67\\
 \cline{2-9}\pagebreak[0]
 &  & $\alpha_1 = 0.50 $ & Gama(0.1 , 0.1) & 0.52 & 0.0012 & 0.1990 & 3.68 & 92.67\\
\nopagebreak
 &  & $\alpha_2 = 1.30 $ & Gama(0.1 , 0.1) & 1.30 & 0.0004 & 0.1169 & 0.15 & 94.33\\
\nopagebreak
 &  & $\beta_1 = 0.60 $ & N(0 , 4) & 0.59 & 0.0018 & 0.1433 & -1.63 & 95.33\\
\nopagebreak
 &  & $\beta_2 = 0.80 $ & N(0 , 4) & 0.79 & 0.0009 & 0.0316 & -1.68 & 94.67\\
\nopagebreak
 &  & $\psi_0 = 0.80 $ & N(0 , 4) & 0.90 & 0.0019 & 0.0156 & 12.94 & 86.33\\
\nopagebreak
 &  & $\psi_1 = 1.20 $ & N(0 , 4) & 1.18 & 0.0022 & 0.0667 & -1.42 & 95.33\\
\nopagebreak
\multirow{-13}{*}{\raggedright\arraybackslash \textbf{$\boldsymbol{C_{500}^{6.3}}$}} & \multirow{-7}{*}{\raggedright\arraybackslash ZI-NHPP-SE-COV} & $\tau = 1.00 $ & Gamma(0.01 , 0.01) & 1.11 & 0.0079 & 0.0747 & 11.34 & 94.33\\
\cmidrule{1-9}\pagebreak[0]
 &  & $\alpha_1 = 2.00 $ & Gama(0.1 , 0.1) & 2.00 & 0.0029 & 1.0966 & 0.23 & 95.00\\
\nopagebreak
 &  & $\alpha_2 = 1.30 $ & Gama(0.1 , 0.1) & 1.30 & 0.0002 & 0.1154 & 0.04 & 96.00\\
\nopagebreak
 &  & $\beta_1 = 0.60 $ & N(0 , 4) & 0.60 & 0.0009 & 0.1315 & 0.06 & 95.33\\
\nopagebreak
 &  & $\beta_2 = 0.80 $ & N(0 , 4) & 0.80 & 0.0004 & 0.0264 & -0.05 & 95.67\\
\nopagebreak
 &  & $\pi = 0.75 $ & Beta(1 , 1) & 0.75 & 0.0003 & 0.0441 & 0.29 & 95.00\\
\nopagebreak
 & \multirow{-6}{*}{\raggedright\arraybackslash ZI-NHPP-SE} & $\tau = 1.00 $ & Gamma(0.01 , 0.01) & 1.03 & 0.0040 & 0.0250 & 3.32 & 97.00\\
 \cline{2-9}\pagebreak[0]
 &  & $\alpha_1 = 2.00 $ & Gama(0.1 , 0.1) & 2.01 & 0.0017 & 1.1034 & 0.42 & 93.00\\
\nopagebreak
 &  & $\alpha_2 = 1.30 $ & Gama(0.1 , 0.1) & 1.30 & 0.0001 & 0.1149 & -0.01 & 94.33\\
\nopagebreak
 &  & $\beta_1 = 0.60 $ & N(0 , 4) & 0.60 & 0.0005 & 0.1330 & -0.31 & 93.33\\
\nopagebreak
 &  & $\beta_2 = 0.80 $ & N(0 , 4) & 0.80 & 0.0002 & 0.0264 & -0.06 & 96.67\\
\nopagebreak
 &  & $\psi_0 = 0.80 $ & N(0 , 4) & 0.82 & 0.0013 & 0.0279 & 2.09 & 93.67\\
\nopagebreak
 &  & $\psi_1 = 1.20 $ & N(0 , 4) & 1.22 & 0.0015 & 0.0775 & 1.74 & 93.67\\
\nopagebreak
\multirow{-13}{*}{\raggedright\arraybackslash \textbf{$\boldsymbol{C_{1000}^{25}}$}} & \multirow{-7}{*}{\raggedright\arraybackslash ZI-NHPP-SE-COV} & $\tau = 1.00 $ & Gamma(0.01 , 0.01) & 1.03 & 0.0029 & 0.0245 & 2.79 & 96.33\\
\bottomrule
\end{tabular*}
\end{table*}

\subsection{Evaluation of Semiparametric Models}\label{subsec2}

In this section, similar to the previous one, semiparametric models will be evaluated concerning their capacity to retrieve the actual parameter values using the same datasets as in the previous section. Following that, a graphical analysis will assess the models' ability to approximate the curve shape of the baseline intensity function. Lastly, the quality of fit between the parametric models and their semiparametric counterparts will be comparatively analyzed using information criterion measures such as WAIC and PSIS-LOO.

To determine the degree of the polynomial, we considered the shape of the curve produced by the Power Law Process, which is monotonous and exhibits smooth behavior. In such conditions, polynomials of degree $d=5$ demonstrate a good ability to approximate the intensity function.

The results shown in Table \ref{tab:tab3} indicate the good performance of the models in recovering the real values of the parameters, even in scenarios with reduced sample sizes. Similar to parametric versions, increasing the average number of expected events leads to smaller relative biases for most of the evaluated parameters. For example, when analyzing the estimated biases for the parameter $\tau$ of the SZI-NHPP-SE model, a reduction of $14.07\%$ is noted in scenario $C_{300}^{6.3}$, decreasing to $7.75\%$ and subsequently to $3.28\%$ in scenarios $C_{300}^{12.5}$ and $C_{300}^{25}$, respectively. In the scenario with the highest number of expected events, $C_{1000}^{25}$, all observed values for biases are less than $3.30\%$. This pattern is also observed for the SZI-NHPP-SE-COV model, as corroborated by the schematic diagrams available at \href{https://alissonccs.shinyapps.io/DASH_SCHEMATIC_DIAGRAMS/}{https://alissonccs.shinyapps.io/DASH\_SCHEMATIC\_DIAGRAMS/}.

\begin{table*}[h]
\caption{Mean, estimated standard error, standard error of estimates, average relative bias (\%) and coverage rate (\%) for the models in scenarios $\boldsymbol{C_{300}^{6.3}}$, $\boldsymbol{C_{300}^{25}}$ and $\boldsymbol{C_{1000}^{25}}$.\label{tab:tab3}}
\tabcolsep=0pt
\begin{tabular*}{\textwidth}{@{\extracolsep{\fill}}llllcccrr@{\extracolsep{\fill}}}
\toprule%
\textbf{Scenarios} &\textbf{Model} & \textbf{Parameter} & \textbf{Prior Dist.} & \textbf{Average Est.} & \textbf{SE} & \textbf{SDE} & \textbf{RB (\%)} & \textbf{CR (\%)}\\
\midrule
 &  & $\beta_1 = 0.60 $ & N(0 , 4) & 0.59 & 0.0047 & 0.0133 & -2.18 & 96.33\\
\nopagebreak
 &  & $\beta_2 = 0.80 $ & N(0 , 4) & 0.78 & 0.0021 & 0.0029 & -2.75 & 94.67\\
\nopagebreak
 &  & $\pi = 0.75 $ & Beta(1 , 1) & 0.76 & 0.0005 & 0.0006 & 1.30 & 91.67\\
\nopagebreak
 & \multirow{-4}{*}{\raggedright\arraybackslash SZI-NHPP-SE} & $\tau = 1.00 $ & Gamma(0.01 , 0.01) & 1.14 & 0.0158 & 0.0941 & 14.07 & 97.00\\
\cline{2-9}\pagebreak[0]
 &  & $\beta_1 = 0.60 $ & N(0 , 4) & 0.59 & 0.0030 & 0.0082 & -0.87 & 95.33\\
\nopagebreak
 &  & $\beta_2 = 0.80 $ & N(0 , 4) & 0.78 & 0.0014 & 0.0018 & -2.11 & 93.00\\
\nopagebreak
 &  & $\psi_0 = 0.80 $ & N(0 , 4) & 0.86 & 0.0030 & 0.0188 & 8.09 & 95.67\\
\nopagebreak
 &  & $\psi_1 = 1.20 $ & N(0 , 4) & 1.15 & 0.0033 & 0.0249 & -4.21 & 95.00\\
\nopagebreak
\multirow{-9}{*}{\raggedright\arraybackslash \textbf{$\boldsymbol{C_{300}^{6.3}}$}} & \multirow{-5}{*}{\raggedright\arraybackslash SZI-NHPP-SE-COV} & $\tau = 1.00 $ & Gamma(0.01 , 0.01) & 1.12 & 0.0110 & 0.0775 & 12.14 & 96.00\\
\cmidrule{1-9}\pagebreak[0]
 &  & $\beta_1 = 0.60 $ & N(0 , 4) & 0.60 & 0.0035 & 0.0723 & 0.39 & 94.00\\
\nopagebreak
 &  & $\beta_2 = 0.80 $ & N(0 , 4) & 0.80 & 0.0015 & 0.0045 & 0.15 & 94.67\\
\nopagebreak
 &  & $\pi = 0.75 $ & Beta(1 , 1) & 0.75 & 0.0004 & 0.0127 & -0.27 & 93.67\\
\nopagebreak
 & \multirow{-4}{*}{\raggedright\arraybackslash SZI-NHPP-SE} & $\tau = 1.00 $ & Gamma(0.01 , 0.01) & 1.08 & 0.0103 & 0.1115 & 7.75 & 93.67\\
\cline{2-9}\pagebreak[0]
 &  & $\beta_1 = 0.60 $ & N(0 , 4) & 0.60 & 0.0019 & 0.0702 & -0.26 & 96.67\\
\nopagebreak
 &  & $\beta_2 = 0.80 $ & N(0 , 4) & 0.80 & 0.0008 & 0.0042 & -0.10 & 93.33\\
\nopagebreak
 &  & $\psi_0 = 0.80 $ & N(0 , 4) & 0.82 & 0.0026 & 0.0255 & 2.19 & 95.00\\
\nopagebreak
 &  & $\psi_1 = 1.20 $ & N(0 , 4) & 1.22 & 0.0030 & 0.1617 & 1.74 & 95.00\\
\nopagebreak
\multirow{-9}{*}{\raggedright\arraybackslash \textbf{$\boldsymbol{C_{300}^{25}}$}} & \multirow{-5}{*}{\raggedright\arraybackslash SZI-NHPP-SE-COV} & $\tau = 1.00 $ & Gamma(0.01 , 0.01) & 1.07 & 0.0070 & 0.0915 & 7.19 & 93.67\\
\cmidrule{1-9}\pagebreak[0]
 &  & $\beta_1 = 0.60 $ & N(0 , 4) & 0.60 & 0.0009 & 0.0676 & -0.14 & 93.67\\
\nopagebreak
 &  & $\beta_2 = 0.80 $ & N(0 , 4) & 0.80 & 0.0004 & 0.0035 & -0.02 & 94.00\\
\nopagebreak
 &  & $\pi = 0.75 $ & Beta(1 , 1) & 0.75 & 0.0003 & 0.0112 & 0.31 & 94.00\\
\nopagebreak
 & \multirow{-4}{*}{\raggedright\arraybackslash SZI-NHPP-SE} & $\tau = 1.00 $ & Gamma(0.01 , 0.01) & 1.03 & 0.0041 & 0.0499 & 3.28 & 97.33\\
\cline{2-9}\pagebreak[0]
 &  & $\beta_1 = 0.60 $ & N(0 , 4) & 0.60 & 0.0005 & 0.0672 & -0.09 & 95.67\\
\nopagebreak
 &  & $\beta_2 = 0.80 $ & N(0 , 4) & 0.80 & 0.0002 & 0.0034 & -0.06 & 96.00\\
\nopagebreak
 &  & $\psi_0 = 0.80 $ & N(0 , 4) & 0.82 & 0.0013 & 0.0083 & 2.18 & 94.67\\
\nopagebreak
 &  & $\psi_1 = 1.20 $ & N(0 , 4) & 1.22 & 0.0016 & 0.1430 & 1.77 & 94.33\\
\nopagebreak
\multirow{-9}{*}{\raggedright\arraybackslash \textbf{$\boldsymbol{C_{1000}^{25}}$}} & \multirow{-5}{*}{\raggedright\arraybackslash SZI-NHPP-SE-COV} & $\tau = 1.00 $ & Gamma(0.01 , 0.01) & 1.03 & 0.0029 & 0.0488 & 2.73 & 97.00\\
\bottomrule
\end{tabular*}
\end{table*}

The curves of the baseline intensity functions estimated by the SZI-NHPP-SE-COV model are shown in Figure \textbf{3}. These figures also display the curves generated by the parametric versions and the curve generated from the real values of the parameters. The curves generated by the other models are available at \href{https://alissonccs.shinyapps.io/DASH_INT_FUNC_SPATIAL_MODEL/}{https://alissonccs.shinyapps.io/DASH\_INT\_FUNC\_SPATIAL\_MODEL/}. Graphical analysis suggests better performance of parametric versions of the models. However, the semiparametric versions present apparently similar results in scenarios with a larger sample size and a higher number of expected events. It can also be noted that the curves tend to concentrate increasingly closer to the real curve as the sample size increases.


\begin{figure*}[h!]%
\centering
\includegraphics[width=0.7\textwidth]{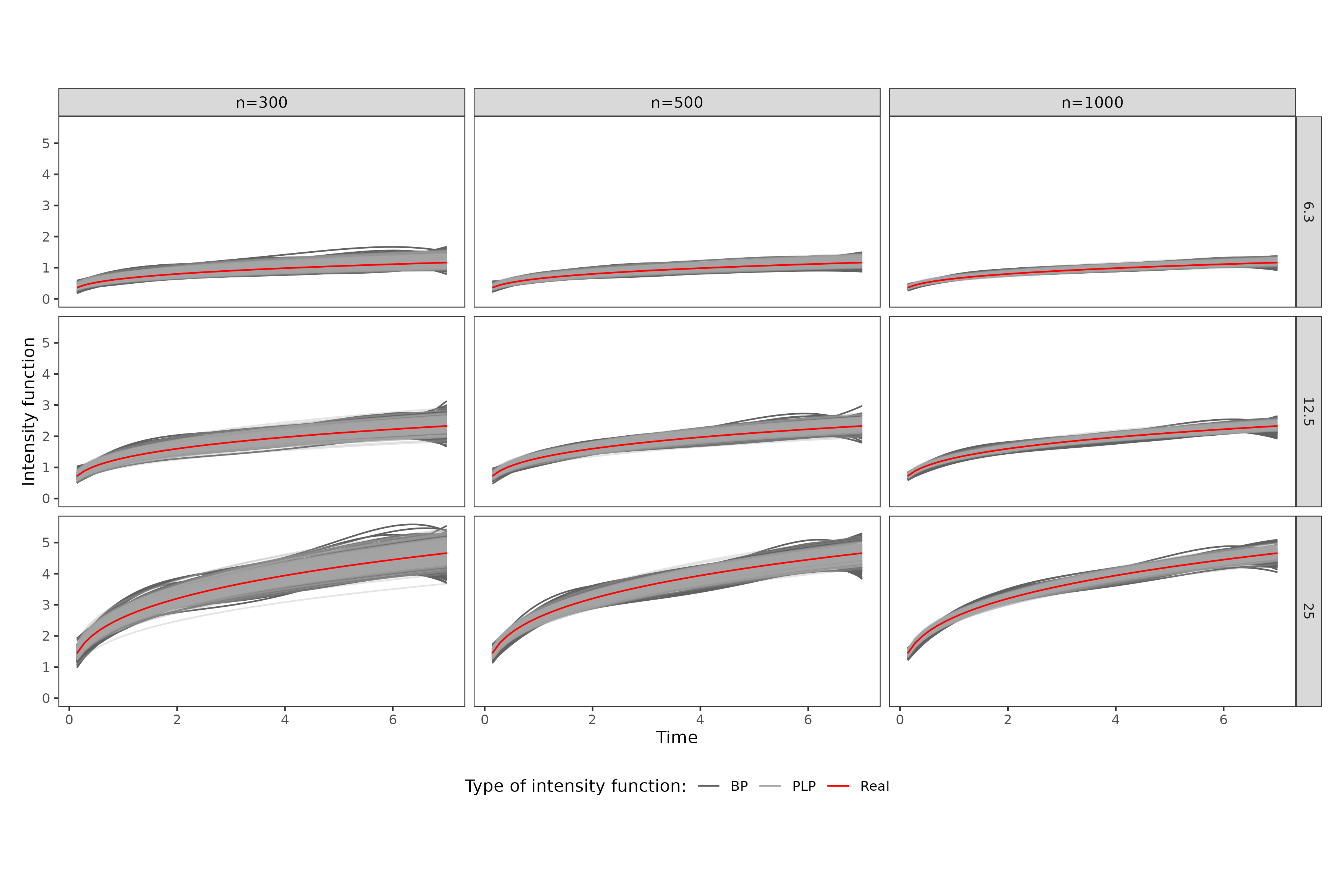}
\caption{Real curve and estimated curves for the baseline intensity function for the ZI-NHPP-SE-COV and SZI-NHPP-SE-COV models, by sample size and expected number of events - Data generating model: PLP}\label{fig:FI2}
\end{figure*}

The results generated for the information criteria indicate that the performance of the ZI-NHPP-SE parametric model is affected by scenarios with a low volume of information. For slightly over half of the total samples generated in Scenario \(C_{300}^{6.3}\), the value of the PSIS-LOO information criterion was lower for the parametric model, accounting for \(53.12\%\) of cases. When evaluating the results of fitting to Scenario \(C_{300}^{25}\) datasets, the percentage of cases in which the parametric model presents itself as the best fit is greater than \(92.00\%\). For the ZI-NHPP-SE-COV model, when considering Scenario \(C_{1000}^{25}\), the percentage of selection of the parametric model reaches \(34.00\%\) of the total number of replications. This result is different from that observed for the ZI-NHPP-SE model and indicates better performance of the semiparametric version, even in a scenario with a high volume of data.

The findings from this section highlight that parametric models generally outperformed semiparametric models across most evaluated scenarios. This outcome was anticipated due to the parametric nature of the generating models. Moreover, the results suggest that the semiparametric models provided satisfactory approximations of the baseline intensity curves.

\subsection{Evaluation of Models Using Non-Monotonic Generator Model}\label{subsec2}

The aim of this section is to assess the performance of the ZI-NHPP-SE and SZI-NHPP-SE models in approximating baseline intensity functions with a more flexible form than the Power Law. We conducted 300 Monte Carlo replications for sample sizes $n=300$, $n=500$, and $n=1000$. The baseline intensity generating function is defined by the expression:
\begin{align} \label{"FI_polinomial"}
\lambda(t)= 1 + b(1+\sin(0.25\pi t)),
\end{align}
where \(b\) is a constant, and \(t\) represents the instant of occurrence of the event.

In determining the polynomial degree for the semiparametric model, we followed the methodology outlined by \citet{osman2012nonparametric}, where the polynomial order is determined as $d=n^{0.4}$. For a sample size of $n=300$, the polynomial degree is set to $d=10$. With larger sample sizes of $n=500$ and $n=1000$, the polynomial degrees become $d=12$ and $d=16$ respectively.

The model's performance will be evaluated by analyzing its ability to recover the true parameter values and through graphical analysis of the estimated baseline intensity functions. Additionally, a sensitivity analysis will be conducted to assess the impact of the polynomial degree on the model fit quality.

\begin{figure*}[h!]%
\centering
\includegraphics[width=0.7\textwidth]{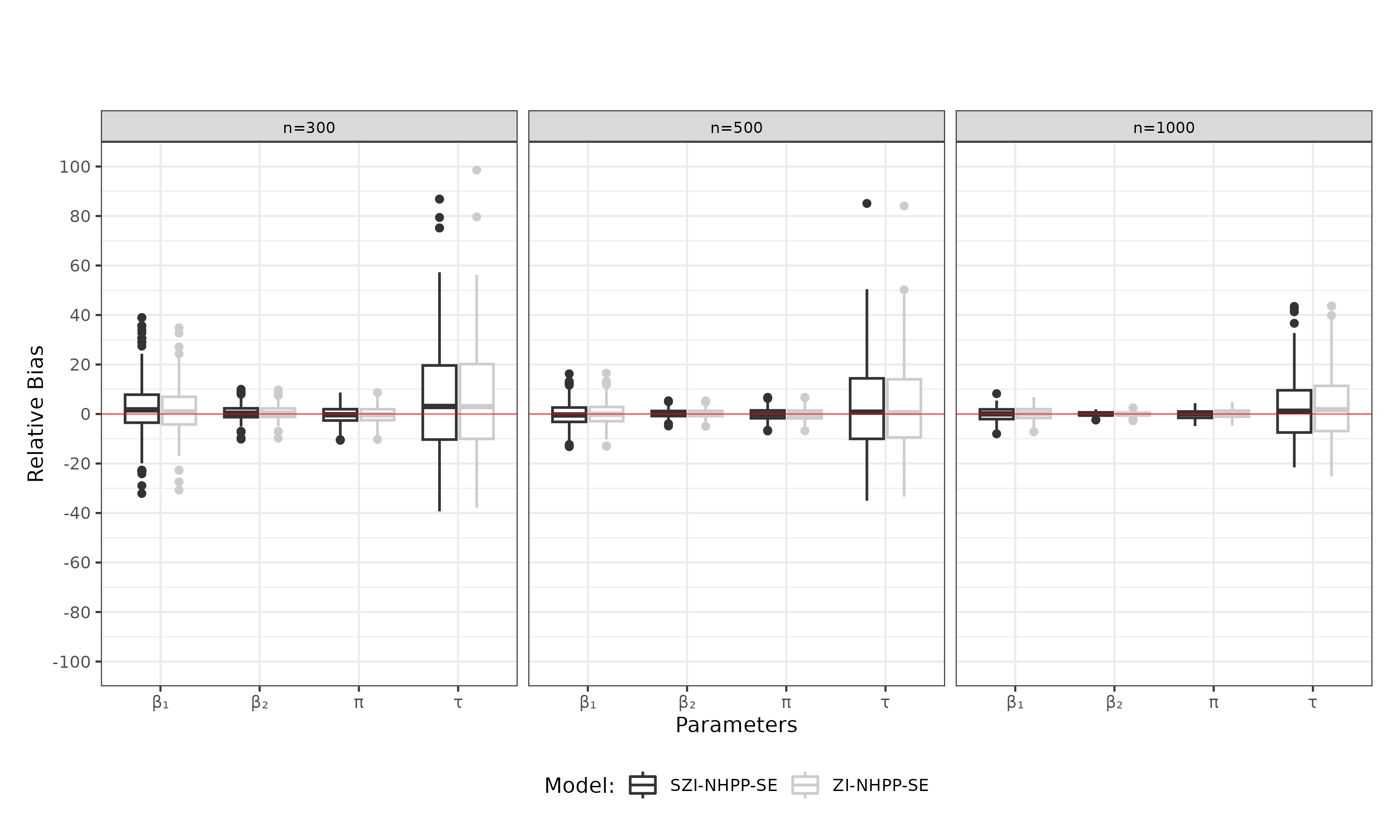}
\caption{Schematic diagrams for the estimated relative biases of the parameters for the ZI-NHPP-SE and SZI-NHPP-SE models, by sample size - Data generating model: Non-monotonic.}\label{fig:BOX_POLI}
\end{figure*}

Figure \ref{fig:BOX_POLI} displays the distribution of relative biases in estimates generated by the ZI-NHPP-SE and SZI-NHPP-SE models. Generally, the biases cluster around zero, with magnitudes mostly below $20.00\%$ for most parameters, except for the spatial precision parameter $\tau$. Analysis of the PSIS-LOO information criterion indicates that, for the datasets under evaluation, the semiparametric model outperforms the parametric model in terms of fit. In a sample of size $n=300$, the semiparametric model demonstrated a superior fit in $100\%$ of comparisons. This percentage decreases to $87.37\%$ and $82.32\%$ when the sample sizes increase to $n=500$ and $n=1000$, respectively.

\begin{figure*}[h!]%
\centering
\includegraphics[width=0.8\textwidth]{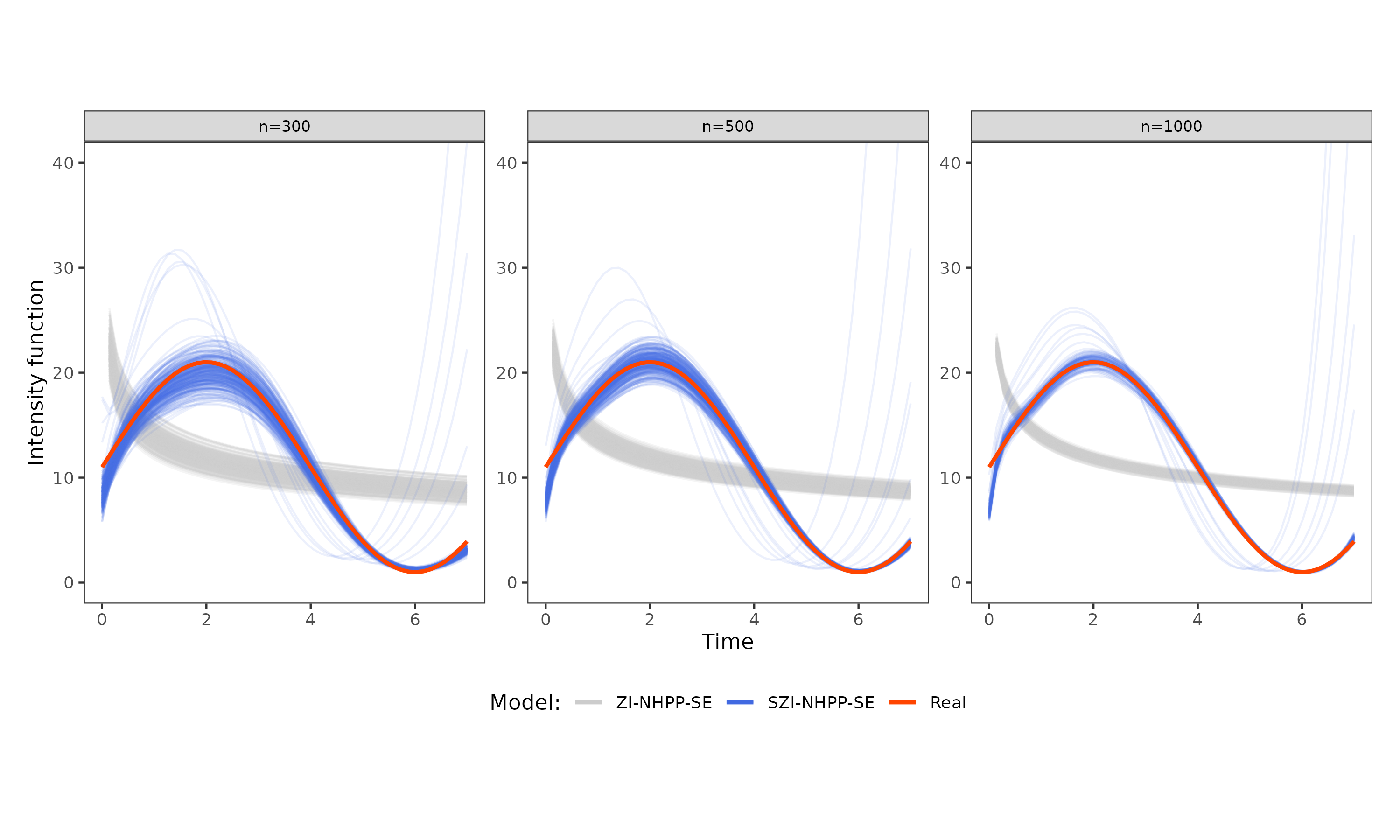}
\caption{Real curve and estimated curves for the baseline intensity function for the ZI-NHPP-SE and SZI-NHPP-SE models, by sample size - Data generating model: Non-monotonic.}\label{fig:FI_SIM_POLI_BP_PLP}
\end{figure*}

The graphs shown in Figure \ref{fig:FI_SIM_POLI_BP_PLP} display, in blue, the baseline intensity curves estimated by the ZI-NHPP-SE and SZI-NHPP-SE models, along with the real curve generated by the polynomial presented in Equation \eqref{"FI_polinomial"}. It can be observed that, for all configurations considered, the curves generated by the SZI-NHPP-SE model cluster around the real curve used to generate the data. Among the 300 replicas, only a small number of curves deviate from the shape of the real curve. As discussed in previous sections, the Power Law process exhibits a monotonous form that can be increasing, decreasing, or constant, making it unsuitable for modeling functions with flexible shapes, as demonstrated in Figure \ref{fig:FI_SIM_POLI_BP_PLP}, in grey.

To evaluate the effect of polynomial degree on the quality of model fit, an analysis was conducted fitting the SZI-NHPP-SE model to the same dataset with sample size \(n=500\), considering polynomial degrees \(d=(5,10,12,15,20,25,30)\). For each fit, the values of the PSIS-LOO and WAIC information criteria were calculated. The results obtained from PSIS-LOO point to the fit with degree \(d=25\) as the best achieved, as it presents the lowest value for the information criteria. This result may be attributed to the complexity of the curve shape of the proposed baseline intensity function. Conversely, the WAIC criterion suggests that \(d=30\) is the best fit. However, due to the robustness of the PSIS-LOO criterion, the result indicated by it will be considered.
The fitting of the ZI-NHPP-SE parametric model yielded a PSIS-LOO criterion value of $-127228.87$, the highest among all fits conducted. This underscores the inadequacy of the Power Law form as an estimated intensity function.

The importance of implementing semiparametric versions of the evaluated models is evidenced by the results of this section. By adopting a baseline intensity function with greater flexibility in the data generation model, the performance of the semiparametric model surpassed that of the parametric model. In addition to demonstrating a good ability to recover the true values of the parameters, the model provided a more accurate approximation of the true shape of the baseline intensity function, being consistently identified in most replicates as the model with the best fit to the analyzed data.

\section{Application}\label{subsec2}

We revisit the data presented in the Database section.
As discussed this study focuses on police recidivism, specifically examining cases where individuals with prior involvement in crimes related to aggravated assault are again documented by the police for engaging in new offenses of a similar nature. Our intention is to assess several aspects concerning the trajectory of committing the crime of bodily, with a primary goal of gathering insights to address the following inquiries: Is there a trend related to the frequency of recidivism? Is the sex covariate a factor associated with the frequency of recidivism? What is the probability of recidivism? Is there an influence of geographic space on the pattern of recidivism?

To perform these analyses, we will fit the models outlined in this study using the \texttt{spnhppzi} package. For each fit, we will generate $2$ parallel chains, each with a size of $2,000$ samples, and discard the initial $1000$ samples from each chain during the warm-up phase, like presented in Section~\ref{subsec2}. Again we used the $\hat R$ statistics to verify convergence.
For the parametric models, the baseline intensity function will be modeled using the Power Law process. On the other hand, for the semiparametric models, we will perform 20 fittings for each model, adjusting the degrees of polynomials from 1 to 20. The PSIS-LOO criterion will guide us in identifying the optimal polynomial degree that offers the best fit for the semiparametric models and in selecting the overall best fit. After selecting the best fit model, we will proceed with an analysis of the results. 

To standardize the types of study units used, municipalities from the Metropolitan Region of Belo Horizonte were selected, which have delineations for weighting areas and exhibit contiguous urban settlement with the municipality of Belo Horizonte. Therefore, including the municipality of Belo Horizonte, five municipalities are covered, totaling 133 weighting areas, as illustrated in Figure \ref{fig:mapa_percent_reincid}. 
As mentioned previously, the dataset includes only the sex covariate. To avoid identifiability problems outlined in the work of \citet{diop2011maximum}, the covariate will be handled in two different ways: either as a factor associated with the intensity function, when fitting the SZI-NHPP-SE model, or as a factor related to the probability of recidivism, when fitting the SZI-NHPP-SE-COV model.

Table \ref{tab:tabresultap1} presents the results obtained from fitting the models. It includes details about the prior distributions employed, the degree of the polynomial utilized for fitting the semiparametric models, and the values computed for the PSIS-LOO information criterion. Moreover, the table displays the estimates generated for each parameter and provides the $95\%$ credible intervals (CI) 
obtained from the estimated posterior distributions.
Upon examining the calculated values of the PSIS-LOO criterion, it becomes evident that, in general, the semiparametric versions of the models exhibit a higher quality of fit compared to their corresponding parametric versions. Furthermore, the manner in which the sex covariate is utilized proves to be another crucial factor affecting the model's fitting quality. Specifically, employing this covariate as a factor linked to the likelihood of recidivism yields a superior fit in contrast to when it is used as a factor associated with the intensity function. This distinction is apparent when comparing the PSIS-LOO criterion values of the ZI-NHPP-SE models ($62158.41$) with those of the ZI-NHPP-SE-COV model ($62091.52$). A similar pattern is observed for their respective semiparametric versions.
The calculated PSIS-LOO values for the SZI-NHPP-SE-COV and ZI-NHPP-SE-COV models suggest that they yielded, respectively, the top two fits among all the models assessed. This outcome affirms that the integration of the zero-inflated data structure, spatial random effects, and the use of the covariate as a factor linked to the likelihood of recidivism leads to the best fit among all models.
\begin{table*}[h]
\caption{Prior Distributions and Fitting Results of the Presented Models. \label{tab:tabresultap1}}
\tabcolsep=0pt
\begin{tabular*}{\textwidth}{@{\extracolsep{\fill}}lclccl@{\extracolsep{\fill}}}
\toprule
\textbf{Model} & \textbf{Parameter} & \textbf{Prior Dist.} & \textbf{Degree} & \textbf{PSIS-LOO} & \textbf{Estimate (C.I)}\\
\midrule\pagebreak[0]
 & $\alpha_1$ & Gamma(0.1 , 0.1) & - &  & 0.00056 (0.00051;0.00061)\\
\nopagebreak
 & $\alpha_2$ & Gamma(0.1 , 0.1) & - &  & 0.95971 (0.94829;0.97076)\\
\nopagebreak
\multirow{-3}{*}{\raggedright\arraybackslash \textbf{NHPP}} & $\beta_1$ & N(0 , 4) & - & \multirow{-3}{*}{\centering\arraybackslash 488816.23} & 0.02575 (-0.00282;0.05514)\\
\cmidrule{1-6}\pagebreak[0]
 & $\alpha_1$ & Gamma(0.1 , 0.1) & - &  & 0.00022 (0.00016;3e-04)\\
\nopagebreak
 & $\alpha_2$ & Gamma(0.1 , 0.1) & - &  & 1.08866 (1.0493;1.12583)\\
\nopagebreak
 & $\beta_1$ & N(0 , 4) & - &  & 0.33091 (0.18932;0.47662)\\
\nopagebreak
\multirow{-4}{*}{\raggedright\arraybackslash \textbf{ZI-NHPP}} & $\pi$ & Beta(1 , 1) & - & \multirow{-4}{*}{\centering\arraybackslash 62163.92} & 0.91232 (0.90784;0.91667)\\
\cmidrule{1-6}\pagebreak[0]
 & $\alpha_1$ & Gamma(0.1 , 0.1) & - &  & 0.00021 (0.00015;0.00029)\\
\nopagebreak
 & $\alpha_2$ & Gamma(0.1 , 0.1) & - &  & 1.08848 (1.04912;1.12642)\\
\nopagebreak
 & $\beta_1$ & N(0 , 4) & - &  & 0.33905 (0.19478;0.48222)\\
\nopagebreak
 & $\pi$ & Beta(1 , 1) & - &  & 0.91061 (0.90595;0.91507)\\
\nopagebreak
\multirow{-5}{*}{\raggedright\arraybackslash \textbf{ZI-NHPP-SE}} & $\tau$ & Gamma(0.01 , 0.1) & - & \multirow{-5}{*}{\centering\arraybackslash 62158.41} & 8.13368 (4.60091;13.01442)\\
\cmidrule{1-6}\pagebreak[0]
 & $\alpha_1$ & Gamma(0.1 , 0.1) & - &  & 0.00029 (0.00022;0.00039)\\
\nopagebreak
 & $\alpha_2$ & Gamma(0.1 , 0.1) & - &  & 1.0909 (1.05204;1.12676)\\
\nopagebreak
 & $\psi_0$ & N(0 , 4) & - &  & 2.88003 (2.75968;3.01683)\\
\nopagebreak
 & $\psi_1$ & N(0 , 4) & - &  & -0.63182 (-0.77388;-0.49697)\\
\nopagebreak
\multirow{-5}{*}{\raggedright\arraybackslash \textbf{ZI-NHPP-SE-COV}} & $\tau$ & Gamma(0.01 , 0.01) & - & \multirow{-5}{*}{\centering\arraybackslash 62091.52} & 1.01647 (0.12998;3.23197)\\
\cmidrule{1-6}\pagebreak[0]
 & $\beta_1$ & N(0 , 4) & 4 &  & 0.31788 (0.17408;0.46071)\\
\nopagebreak
 & $\pi$ & Beta(1 , 1) & 4 &  & 0.91092 (0.90593;0.91544)\\
\nopagebreak
\multirow{-3}{*}{\raggedright\arraybackslash \textbf{SZI-NHPP-SE}} & $\tau$ & Gamma(0.01 , 0.01) & 4 & \multirow{-3}{*}{\centering\arraybackslash 62141.75} & 8.1994 (4.70543;12.8898)\\
\cmidrule{1-6}\pagebreak[0]
 & $\psi_0$ & N(0 , 4) & 4 &  & 2.88215 (2.7532;3.01343)\\
\nopagebreak
 & $\psi_1$ & N(0 , 4) & 4 &  & -0.6304 (-0.7731;-0.49781)\\
\nopagebreak
\multirow{-3}{*}{\raggedright\arraybackslash \textbf{SZI-NHPP-SE-COV}} & $\tau$ & Gamma(0.01 , 0.01) & 4 & \multirow{-3}{*}{\centering\arraybackslash 62075.29} & 0.81425 (0.07925;2.57954)\\
\bottomrule
\end{tabular*}
\end{table*} 

\begin{table*}[h]
\caption{Prior Distributions and Fitting Results of the Presented Models. \label{tab:tabresultap1}}
\tabcolsep=0pt
\begin{tabular*}{\textwidth}{@{\extracolsep{\fill}}lclccl@{\extracolsep{\fill}}}
\toprule
\textbf{Model} & \textbf{Parameter} & \textbf{Prior Dist.} & \textbf{Degree} & \textbf{PSIS-LOO} & \textbf{Estimate (C.I)}\\
\midrule\pagebreak[0]
 & $\alpha_1$ & Gamma(0.1 , 0.1) & - &  & 0.00056 (0.00051 ; 0.00061)\\
\nopagebreak
 & $\alpha_2$ & Gamma(0.1 , 0.1) & - &  & 0.95975 (0.94806 ; 0.97063)\\
\nopagebreak
\multirow{-3}{*}{\raggedright\arraybackslash \textbf{NHPP}} & $\beta_1$ & N(0 , 4) & - & \multirow{-3}{*}{\centering\arraybackslash 488816.21} & 0.02623 (-0.02568 ; 0.05533)\\
\cmidrule{1-6}\pagebreak[0]
 & $\alpha_1$ & Gamma(0.1 , 0.1) & - &  & 0.00021 (0.00015 ; 0.00029)\\
\nopagebreak
 & $\alpha_2$ & Gamma(0.1 , 0.1) & - &  & 1.09153 (1.05359 ; 1.13019)\\
\nopagebreak
 & $\beta_1$ & N(0 , 4) & - &  & 0.33819 (0.18143 ; 0.48468)\\
\nopagebreak
\multirow{-4}{*}{\raggedright\arraybackslash \textbf{ZI-NHPP}} & $\pi$ & Beta(1 , 1) & - & \multirow{-4}{*}{\centering\arraybackslash 62164.29} & 0.91212 (0.90765 ; 0.91645)\\
\cmidrule{1-6}\pagebreak[0]
 & $\alpha_1$ & Gamma(0.1 , 0.1) & - &  & 0.00021 (0.00015 ; 0.00029)\\
\nopagebreak
 & $\alpha_2$ & Gamma(0.1 , 0.1) & - &  & 1.09040 (1.05190 ; 1.12694)\\
\nopagebreak
 & $\beta_1$ & N(0 , 4) & - &  & 0.34014 (0.19637 ; 0.48418)\\
\nopagebreak
 & $\pi$ & Beta(1 , 1) & - &  & 0.91166 (0.90693 ; 0.91641)\\
\nopagebreak
\multirow{-5}{*}{\raggedright\arraybackslash \textbf{ZI-NHPP-SE}} & $\tau$ & Gamma(0.01 , 0.1) & - & \multirow{-5}{*}{\centering\arraybackslash 62159.86} & 87.08068 (13.82129 ; 305.12514)\\
\cmidrule{1-6}\pagebreak[0]
 & $\alpha_1$ & Gamma(0.1 , 0.1) & - &  & 0.00029 (0.00022 ; 0.00039)\\
\nopagebreak
 & $\alpha_2$ & Gamma(0.1 , 0.1) & - &  & 1.09000 (1.05364 ; 1.12579)\\
\nopagebreak
 & $\psi_0$ & N(0 , 4) & - &  & 2.89494 (2.77436 ; 3.01101)\\
\nopagebreak
 & $\psi_1$ & N(0 , 4) & - &  & -0.64853 (-0.77015 ; -0.52146)\\
\nopagebreak
\multirow{-5}{*}{\raggedright\arraybackslash \textbf{ZI-NHPP-SE-COV}} & $\tau$ & Gamma(0.01 , 0.01) & - & \multirow{-5}{*}{\centering\arraybackslash 62091.52} & 3.6530 (0.00047 ; 25.26446)\\
\cmidrule{1-6}\pagebreak[0]
 & $\beta_1$ & N(0 , 4) & 4 &  & 0.33759 (0.18601 ; 0.48063)\\
\nopagebreak
 & $\pi$ & Beta(1 , 1) & 4 &  & 0.91185 (0.90734 ; 0.91613)\\
\nopagebreak
\multirow{-3}{*}{\raggedright\arraybackslash \textbf{SZI-NHPP-SE}} & $\tau$ & Gamma(0.01 , 0.01) & 4 & \multirow{-3}{*}{\centering\arraybackslash 62143.45} & 70.05179 (15.16055 ; 233.40856)\\
\cmidrule{1-6}\pagebreak[0]
 & $\psi_0$ & N(0 , 4) & 4 &  & 2.89855 (2.78463 ; 3.02005)\\
\nopagebreak
 & $\psi_1$ & N(0 , 4) & 4 &  & -0.65107 (-0.77475 ; -0.51951)\\
\nopagebreak
\multirow{-3}{*}{\raggedright\arraybackslash \textbf{SZI-NHPP-SE-COV}} & $\tau$ & Gamma(0.01 , 0.01) & 4 & \multirow{-3}{*}{\centering\arraybackslash 62075.69} & 14.32634 (0.00625 ; 99.47547)\\
\bottomrule
\end{tabular*}
\end{table*}

\begin{figure}[h]
\centering
\begin{minipage}{0.48\textwidth}
\centering
\includegraphics[width=\textwidth]{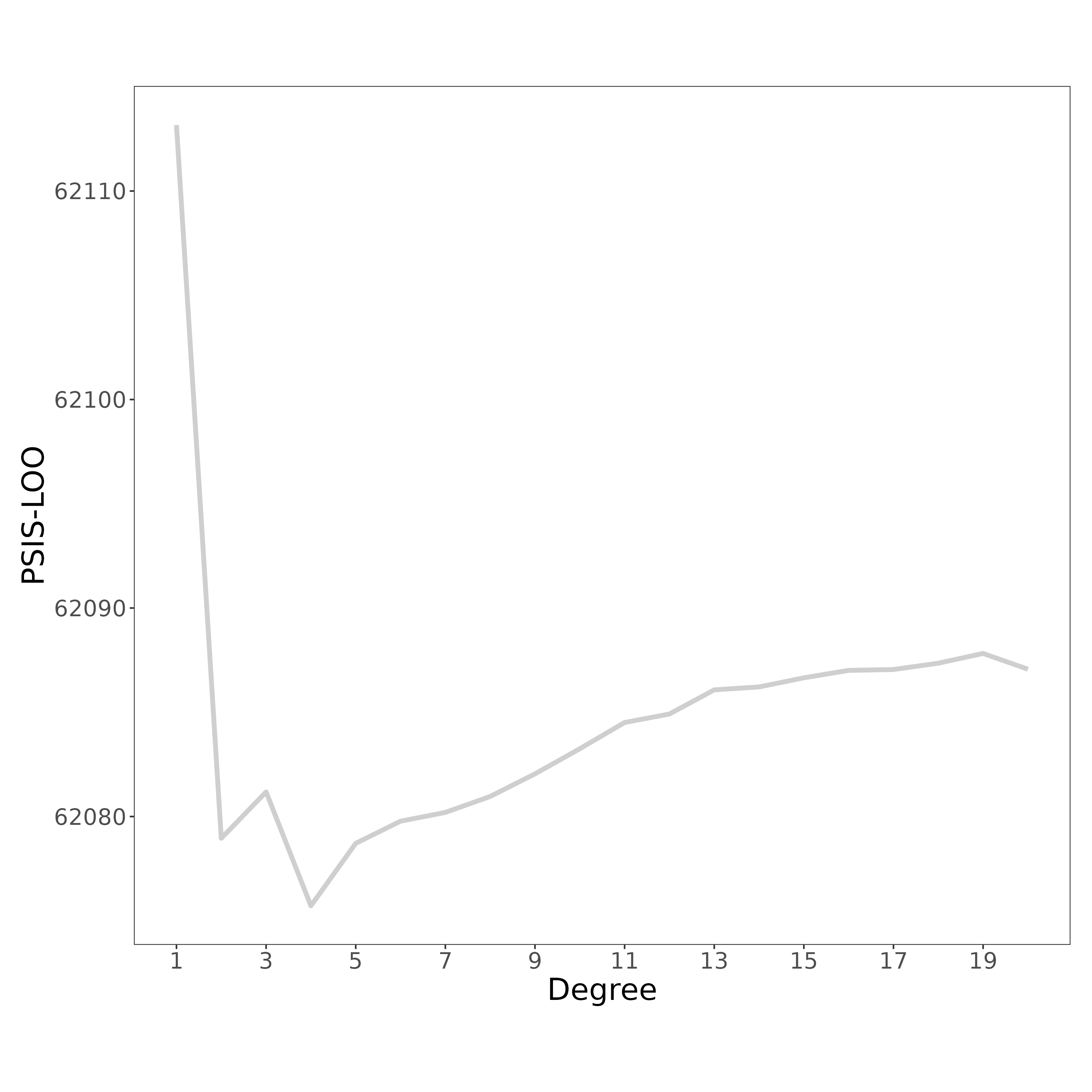}
\caption{PSIS-LOO information criterion per degree of the polynomial for the SZI-NHPP-SE-COV model;}
\label{fig:PSIS-LOO_grau}
\end{minipage}\hfill
\begin{minipage}{0.48\textwidth}
\centering
\includegraphics[width=\textwidth]{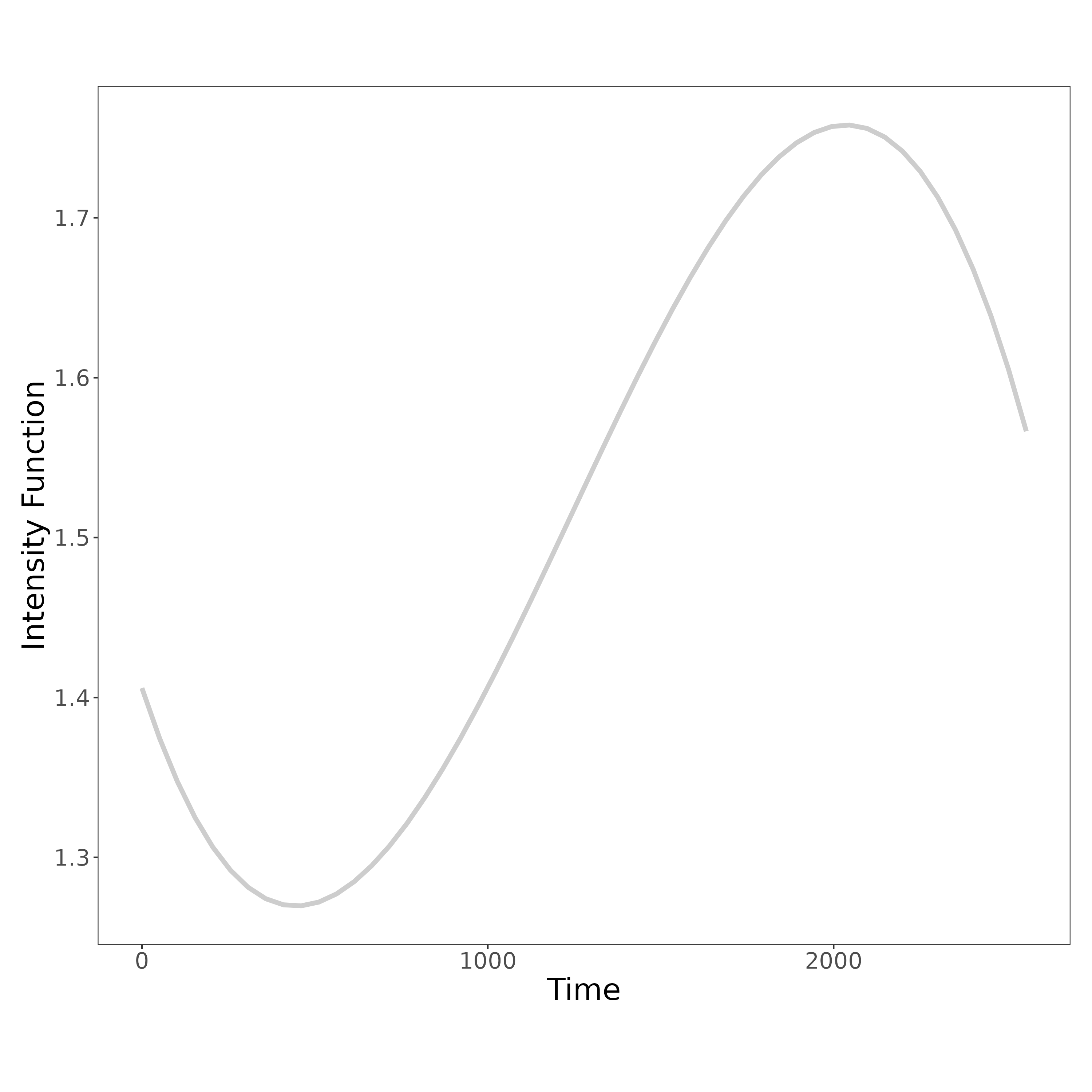}
\caption{Intensity function estimated based on the SZI-NHPP-SE-COV model.}
\label{fig:FI_SZI-NHPP-SE-COV}
\end{minipage}
\end{figure}

As previously mentioned, the degree of the polynomial considered for the SZI-NHPP-SE-COV model was determined through sensitivity analysis. The PSIS-LOO criterion was used to select the optimal degree for achieving the best fit. Analysis of Figure \ref{fig:PSIS-LOO_grau} suggests that, for the SZI-NHPP-SE-COV model, the best fit is obtained with a polynomial of degree $d=4$.


The results provide crucial insights into addressing the issues surrounding criminal recidivism. Regarding the first question concerning trends in the frequency of recidivism for aggravated assault cases, the intensity function estimated by the SZI-NHPP-SE-COV model suggests an oscillatory pattern over the analysis period. Initially, there's a decline in the intensity function, followed by a period of rapid growth. Towards the end, a downward trend is observed until the end of the monitoring period (see Figure \ref{fig:FI_SZI-NHPP-SE-COV}). Studies such as \citet{petersilia1980criminal} delve into the development of criminal careers, aiming to identify behaviors characterizing different phases like early onset, persistent offending and desistance of criminal activities.

Regarding the identification of associated factors, for the SZI-NHPP-SE-COV model, the covariates can influence the probability of recidivism for aggravated assault cases, with their coefficients interpreted through the odds ratio. The results indicate that the probability of non-recidivism for aggravated assault crimes is lower for men, accounting for approximately \(52.0\%\) (\(95\%\) CI: [46.0\%, 59.0\%]) compared to women (see Table \ref{tab:tabresultap1}). Otherwise, the probability of recidivism for men is about 1.48 times higher than for women.

In the SZI-NHPP-SE model, the covariate is linked to the intensity function, and its coefficient can be interpreted as the ratio of intensity functions. The results indicate that the rate of aggravated assault recidivism for men is approximately $40.15\%$ higher than for women (see Table \ref{tab:tabresultap1}). In this model, $\pi$ is assumed constant and represents the estimated probability of non-recidivism for an individual. The estimated value for $\pi$ is approximately $0.91$, which closely aligns with the actual proportion of individuals without recidivism ($0.93$).

\begin{figure}[h]
\normalsize
\centering
    \includegraphics[width=0.5\textwidth]{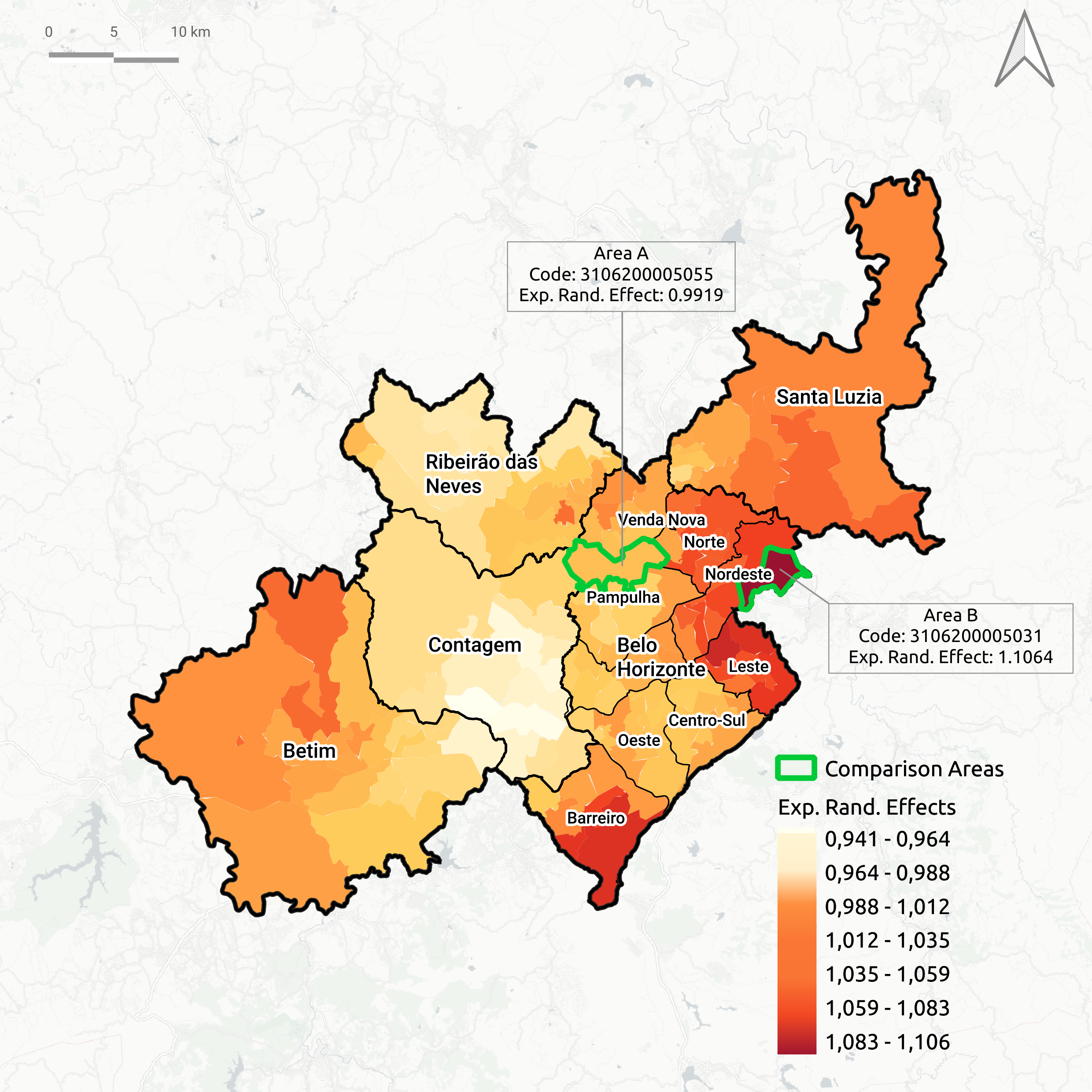}
    \caption{Exponential of the estimated random effects for the SZI-NHPP-SE-COV model, by weighting area.}
    \label{fig:mapa_efeito_aleat_SZI-NHPP-SE-COV}
\end{figure}

Regarding the influence of geographic space, the results obtained from the SZI-NHPP-SE-COV model highlight the spatial disparity concerning the likelihood of criminal recidivism across the analyzed regions. The map depicted in Figure \ref{fig:mapa_efeito_aleat_SZI-NHPP-SE-COV} illustrates that the distribution of random effects exhibits a similar pattern to those observed in the descriptive analysis. The values were exponentiated to facilitate the comparative analysis of the results. Areas in close proximity to the Central and Pampulha regions exhibit lower values for random effects. As one moves away from this central core, the effects tend to increase, particularly in the peripheral regions where they become more pronounced. The analysis of the entire study area indicates a certain level of stability between the central region of Belo Horizonte and parts of the municipality of Betim. However, the westernmost region of Betim tends to have higher rates of random effects. Furthermore, a pattern of escalating effects is noticeable from the central region of Belo Horizonte extending towards the southern region of the municipality of Santa Luzia.

The results presented in Figure \ref{fig:mapa_efeito_aleat_SZI-NHPP-SE-COV} can be interpreted based on the ratio of the intensity functions \citep{cook2007statistical}. Considering areas A and B highlighted on the map, the estimated random effect for area A was $\hat{\omega}_A=0.9919$, while for area B it was $\hat{\omega}_B=1.1064$. The ratio between these values indicates that the recidivism rate for a man who commits aggravated assault in Region B is $11.54\%$ higher than that observed in Region A. In practical terms, this means that if a men in Region A it is expected that a recidivist will commit 10 aggravated assaults over a given period, in Region B, approximately 11.15 crimes would be expected during the same period.

Based on the presented results, it is evident that the models proposed in this paper not only exhibited superior performance compared to traditional models, resulting in more accurate fitting, but also contributed to a deeper understanding of the studied phenomenon. These models incorporate metrics that are typically overlooked by traditional approaches, thereby providing a more comprehensive analysis.

\section{Conclusion}\label{subsec2}

This study introduced novel models for analyzing recurrent events data utilizing counting methods, extending the Non-Homogeneous Poisson Process to address zero-inflated data and spatial autocorrelation. The inclusion of spatial autocorrelation not only mitigates issues of autocorrelation and heteroskedasticity but also enables localized analysis, revealing spatial patterns in event recurrence frequency. Furthermore, the implementation of semiparametric models using Bernstein polynomials expands the applicability of these models in real-world scenarios, particularly in cases with intricate baseline intensity functions.

The simulation study demonstrated that the models are adept at producing precise estimates of the examined parameters, effectively capturing the true values of these parameters, particularly in scenarios with larger datasets. Moreover, semiparametric models proved to be more appropriate than parametric models when dealing with baseline intensity functions of greater flexibility. These findings underscore the significance of semiparametric approaches in handling the diverse patterns encountered in real-world datasets.

The application of the models to the dataset on criminal recidivism in the analyzed region confirms their capability to yield crucial insights into the studied phenomenon, incorporating metrics often overlooked by traditional models. The results underscore spatial disparities in the frequency of criminal recidivism within Belo Horizonte. Peripheral areas exhibit higher recidivism frequencies during the observation period compared to central regions. Notably, the western region of Betim and Santa Luzia stand out with the highest recidivism rates among the municipalities. The sex covariate emerged as a significant factor influencing recidivism frequency, with men showing a $40.15\%$ higher propensity for recidivism in aggravated assault crimes than women. When analyzed as a factor related to recidivism probability, men exhibit a \(1.48\) times higher likelihood of recidivism compared to women. Regarding recidivism frequency trends, parametric models indicate a growth pattern during the evaluated period. In contrast, semiparametric models offer a nuanced analysis, revealing fluctuations in the intensity function throughout the study period.

The development of the \texttt{spnhppzi} package, available at \href{https://github.com/alissonccs/spnhppzi}{https://github.com/alissonccs/spnhppzi}, represents a significant contribution of this paper, both for generating simulated data and fitting the proposed models. However, due to the computational demands of the simulation studies, the exploration of scenarios incorporating other relevant aspects for model evaluation was restricted. Priority was given to scenarios resembling the characteristics observed in the dataset on criminal recidivism. For instance, evaluating model performance in scenarios with non-increasing baseline intensity functions or scenarios including terminal events could yield valuable insights into the behavior of the proposed models. Such an approach would enhance the generalizability of results to various real-world situations. This underscores the importance of future investigations addressing these scenarios, thereby expanding the applicability and robustness of the developed models.

The dataset available for the study of criminal recidivism in the Metropolitan Region of Belo Horizonte includes information on various types of crime, beyond aggravated assault alone. Given this breadth, it is proposed to apply the methodology developed for analyzing other criminal natures as well, and to evaluate patterns of escalation in the severity of crimes committed. To address this issue, there is a proposal to expand the developed models to analyze data from concurrent events or data representing multiple states. This expansion aims to provide a more comprehensive understanding of the dynamics of criminal behavior and the factors influencing the severity and progression of criminal acts.

An important aspect to consider in model development is the assessment of spatial heterogeneity effects on the probability of recidivism. We propose including a random effects structure in the expression presented in \eqref{"logistc"} to address this aspect. Another proposal involves including random effects representing different criminal natures in the dataset, which would enable evaluation of whether susceptibility to recidivism is related to the type of crime \citep{kim2020analysis}. Additionally, there is interest in using the developed models as tools for assessing the risk of recidivism. For example, the Federal District Security Secretariat has a dataset on the recurrence of domestic crimes, including femicides. The objective is to assess the probability of a specific individual committing a femicide, considering various demographic, socioeconomic characteristics, and criminal history. It is crucial to evaluate the predictive power of the models presented for this purpose. This process would not only aid in understanding criminal trajectories but also contribute to developing more effective preventive strategies to protect potential victims.

\section*{Acknowledgements}
Marcos O. Prates acknowledge the research grants obtained from CNPq-Brazil (309186/2021-8), FAPEMIG (APQ-01837-22) and CAPES, respectively for partial financial support. Bráulio F. Silva thank FAPEMIG grant APQ-02325-18 for partial financial support.





\bibliographystyle{plainnat}
\bibliography{reference}

\begin{thebibliography}{58}
\providecommand{\natexlab}[1]{#1}
\providecommand{\url}[1]{\texttt{#1}}
\expandafter\ifx\csname urlstyle\endcsname\relax
  \providecommand{\doi}[1]{doi: #1}\else
  \providecommand{\doi}{doi: \begingroup \urlstyle{rm}\Url}\fi

\bibitem[Adekpedjou and Dabo-Niang(2021)]{adekpedjou2021semiparametric}
Akim Adekpedjou and Sophie Dabo-Niang.
\newblock Semiparametric estimation with spatially correlated recurrent events.
\newblock \emph{Scandinavian Journal of Statistics}, 48\penalty0 (4):\penalty0
  1097--1126, 2021.

\bibitem[Alder and Wainwright(1959)]{alder1959studies}
Berni~J Alder and Thomas~Everett Wainwright.
\newblock Studies in molecular dynamics. {I}. {G}eneral method.
\newblock \emph{The Journal of Chemical Physics}, 31\penalty0 (2):\penalty0
  459--466, 1959.

\bibitem[Andersen and Gill(1982)]{andersen1982cox}
P.~K. Andersen and R.~D. Gill.
\newblock Cox's {R}egression {M}odel for {C}ounting {P}rocesses: {A} {L}arge
  {S}ample {S}tudy.
\newblock \emph{The Annals of Statistics}, 10\penalty0 (4):\penalty0
  1100--1120, 1982.
\newblock ISSN 00905364.
\newblock URL \url{http://www.jstor.org/stable/2240714}.

\bibitem[Andr{\'e}s-Pueyo et~al.(2018)Andr{\'e}s-Pueyo, Arbach-Lucioni, and
  Redondo]{andres2018riscanvi}
Antonio Andr{\'e}s-Pueyo, Karin Arbach-Lucioni, and Santiago Redondo.
\newblock The {R}is{C}anvi: a new tool for assessing risk for violence in
  prison and recidivism.
\newblock \emph{Handbook of recidivism risk/needs assessment tools}, pages
  255--268, 2018.

\bibitem[Assun{\c{c}}{\~a}o et~al.(2002)Assun{\c{c}}{\~a}o, Potter, and
  Cavenaghi]{assunccao2002bayesian}
Renato~M Assun{\c{c}}{\~a}o, Joseph~E Potter, and Suzana~M Cavenaghi.
\newblock A {B}ayesian space varying parameter model applied to estimating
  fertility schedules.
\newblock \emph{Statistics in Medicine}, 21\penalty0 (14):\penalty0 2057--2075,
  2002.

\bibitem[Banerjee et~al.(2003{\natexlab{a}})Banerjee, Carlin, and
  Gelfand]{banerjee2003hierarchical}
Sudipto Banerjee, Bradley~P Carlin, and Alan~E Gelfand.
\newblock \emph{Hierarchical modeling and analysis for spatial data}.
\newblock Chapman and Hall/CRC, 2003{\natexlab{a}}.

\bibitem[Banerjee et~al.(2003{\natexlab{b}})Banerjee, Wall, and
  Carlin]{banerjee2003frailty}
Sudipto Banerjee, Melanie~M Wall, and Bradley~P Carlin.
\newblock Frailty modeling for spatially correlated survival data, with
  application to infant mortality in {M}innesota.
\newblock \emph{Biostatistics}, 4\penalty0 (1):\penalty0 123--142,
  2003{\natexlab{b}}.

\bibitem[Bao et~al.(2019)Bao, Cancho, Louzada, and Suzuki]{bao2019semi}
Yiqi Bao, Vicente~G Cancho, Francisco Louzada, and Adriano~K Suzuki.
\newblock Semi-{P}arametric {C}ure {R}ate {P}roportional {O}dds {M}odels with
  {S}patial {F}railties for {I}nterval-{C}ensored {D}ata.
\newblock \emph{Advances in Data Science and Adaptive Analysis}, 11\penalty0 (3
  \& 4):\penalty0 1--32, 2019.

\bibitem[Bernstein(1912)]{bernstein1912demo}
Serge Bernstein.
\newblock Démonstration du théorème de {W}eierstrass fondée sur le calcul
  des probabilités.
\newblock \emph{Communications de la Société {M}athématique de Kharkow},
  13\penalty0 (1):\penalty0 1--2, 1912.

\bibitem[Besag et~al.(1991)Besag, York, and Molli{\'e}]{besag1991bayesian}
Julian Besag, Jeremy York, and Annie Molli{\'e}.
\newblock Bayesian image restoration, with two applications in spatial
  statistics.
\newblock \emph{Annals of the Institute of Statistical Mathematics},
  43\penalty0 (1):\penalty0 1--20, 1991.

\bibitem[Blumstein and Blumstein(1986)]{blumstein1986criminal}
Alfred Blumstein and Alfred Blumstein.
\newblock \emph{Criminal careers and career criminals}, volume~1.
\newblock National Academy Press Washington, DC, 1986.

\bibitem[Blumstein et~al.(2010)Blumstein, Cohen, Piquero, and
  Visher]{blumstein2010linking}
Alfred Blumstein, Jacqueline Cohen, Alex~R Piquero, and Christy~A Visher.
\newblock Linking the crime and arrest processes to measure variations in
  individual arrest risk per crime (q).
\newblock \emph{Journal of Quantitative Criminology}, 26:\penalty0 533--548,
  2010.

\bibitem[Carnicer and Pena(1993)]{carnicer1993shape}
Jes{\'u}s~M Carnicer and Juan~Manuel Pena.
\newblock Shape preserving representations and optimality of the {B}ernstein
  basis.
\newblock \emph{Advances in Computational Mathematics}, 1\penalty0
  (2):\penalty0 173--196, 1993.

\bibitem[Cook et~al.(2007)Cook, Lawless, et~al.]{cook2007statistical}
Richard~J Cook, Jerald~F Lawless, et~al.
\newblock \emph{The statistical analysis of recurrent events}.
\newblock Springer, New York, 2007.

\bibitem[Cooner et~al.(2006)Cooner, Banerjee, and McBean]{cooner2006modelling}
Freda Cooner, Sudipto Banerjee, and A~Marshall McBean.
\newblock Modelling geographically referenced survival data with a cure
  fraction.
\newblock \emph{Statistical Methods in Medical Research}, 15\penalty0
  (4):\penalty0 307--324, 2006.

\bibitem[Copas and Heydari(2008)]{Copas2008}
J.~Copas and Fakhreddin Heydari.
\newblock Estimating the risk of reoffending by using exponential mixture
  models.
\newblock \emph{Journal of the Royal Statistical Society: Series A (Statistics
  in Society)}, 160:\penalty0 237 -- 252, 10 2008.
\newblock \doi{10.1111/1467-985X.00059}.

\bibitem[da~Silva(2019)]{Rumenickda2019modelos}
Rumenick~Pereira da~Silva.
\newblock \emph{Modelos semiparam{\'e}tricos para an{\'a}lise de eventos
  recorrentes}.
\newblock PhD thesis, Programa de {P}ós-{G}raduação em {E}statística,
  Universidade Federal de Minas Gerais, 2019.

\bibitem[DeLisi(2005)]{delisi2005career}
Matt DeLisi.
\newblock \emph{Career criminals in society}.
\newblock Sage Publications, 2005.

\bibitem[Demarqui et~al.(2012)Demarqui, Loschi, Dey, and
  Colosimo]{DEMARQUI2012728}
Fabio~N. Demarqui, Rosangela~H. Loschi, Dipak~K. Dey, and Enrico~A. Colosimo.
\newblock A class of dynamic piecewise exponential models with random time
  grid.
\newblock \emph{Journal of Statistical Planning and Inference}, 142\penalty0
  (3):\penalty0 728--742, 2012.
\newblock ISSN 0378-3758.
\newblock \doi{https://doi.org/10.1016/j.jspi.2011.09.006}.
\newblock URL
  \url{https://www.sciencedirect.com/science/article/pii/S0378375811003521}.

\bibitem[DEPEN(2022)]{gappe2022reincidencia}
DEPEN.
\newblock Reincidência {C}riminal no {B}rasil.
\newblock Relatório técnico, GAPPE e DEPEN.
  https://www.gov.br/senappen/pt-br/assuntos/noticias/depen-divulga-relatorio-previo-de-estudo-inedito-sobre-reincidencia-criminal-no-brasil/reincidencia-criminal-no-brasil-2022.pdf/@@download/file.,
  2022.
\newblock URL
  \url{https://www.gov.br/senappen/pt-br/assuntos/noticias/depen-divulga-relatorio-previo-de-estudo-inedito-sobre-reincidencia-criminal-no-brasil/reincidencia-criminal-no-brasil-2022.pdf/@@download/file}.

\bibitem[Diop et~al.(2011)Diop, Diop, and Dupuy]{diop2011maximum}
Aba Diop, Aliou Diop, and Jean-Fran{\c{c}}ois Dupuy.
\newblock {Maximum likelihood estimation in the logistic regression model with
  a cure fraction}.
\newblock \emph{Electronic Journal of Statistics}, 5\penalty0 (5):\penalty0 460
  -- 483, 2011.
\newblock \doi{10.1214/11-EJS616}.
\newblock URL \url{https://doi.org/10.1214/11-EJS616}.

\bibitem[Farouki(2012)]{farouki2012bernstein}
Rida~T Farouki.
\newblock The {B}ernstein polynomial basis: A centennial retrospective.
\newblock \emph{Computer Aided Geometric Design}, 29\penalty0 (6):\penalty0
  379--419, 2012.

\bibitem[Gelman et~al.(2014)Gelman, Hwang, and
  Vehtari]{gelman2014understanding}
Andrew Gelman, Jessica Hwang, and Aki Vehtari.
\newblock Understanding predictive information criteria for {B}ayesian models.
\newblock \emph{Statistics and Computing}, 24\penalty0 (6):\penalty0 997--1016,
  2014.

\bibitem[Harries(1974)]{harries1974geography}
Keith~D Harries.
\newblock \emph{The geography of crime and justice}.
\newblock McGraw-Hill New York, 1974.

\bibitem[Ionides(2008)]{ionides2008truncated}
Edward~L Ionides.
\newblock Truncated importance sampling.
\newblock \emph{Journal of Computational and Graphical Statistics}, 17\penalty0
  (2):\penalty0 295--311, 2008.

\bibitem[IPEA(2015)]{ipea2015reincidencia}
IPEA.
\newblock Reincidência {C}riminal no {B}rasil.
\newblock Relatório técnico, Instituto de Pesquisa Econômica Aplicada
  (IPEA), https://repositorio.ipea.gov.br/handle/11058/7510, Brasília -DF,
  2015.
\newblock URL \url{https://repositorio.ipea.gov.br/handle/11058/7510}.

\bibitem[Jahn-Eimermacher et~al.(2015)Jahn-Eimermacher, Ingel, Ozga, Preussler,
  and Binder]{jahn2015simulating}
Antje Jahn-Eimermacher, Katharina Ingel, Ann-Kathrin Ozga, Stella Preussler,
  and Harald Binder.
\newblock Simulating recurrent event data with hazard functions defined on a
  total time scale.
\newblock \emph{BMC Medical Research Methodology}, 15\penalty0 (16):\penalty0
  1--9, 2015.

\bibitem[Kazemian et~al.(2009)Kazemian, Farrington, and
  Le~Blanc]{kazemian2009can}
Lila Kazemian, David~P Farrington, and Marc Le~Blanc.
\newblock Can we make accurate long-term predictions about patterns of
  de-escalation in offending behavior?
\newblock \emph{Journal of Youth and Adolescence}, 38\penalty0 (3):\penalty0
  384--400, 2009.

\bibitem[Kim and Kim(2020)]{kim2020analysis}
Taeun Kim and Yang-Jin Kim.
\newblock Analysis of bivariate recurrent event data with zero inflation.
\newblock \emph{Communications for Statistical Applications and Methods},
  27\penalty0 (1):\penalty0 37--46, 2020.

\bibitem[Kleinbaum et~al.(2012)Kleinbaum, Klein, et~al.]{kleinbaum2012survival}
David~G Kleinbaum, Mitchel Klein, et~al.
\newblock \emph{Survival analysis: a self-learning text}, volume~3.
\newblock Springer, New York, 2012.

\bibitem[Koopman et~al.(2009)Koopman, Shephard, and Creal]{koopman2009testing}
Siem~Jan Koopman, Neil Shephard, and Drew Creal.
\newblock Testing the assumptions behind importance sampling.
\newblock \emph{Journal of Econometrics}, 149\penalty0 (1):\penalty0 2--11,
  2009.

\bibitem[Lambert(1992)]{lambert1992zero}
Diane Lambert.
\newblock Zero-inflated {P}oisson regression, with an application to defects in
  manufacturing.
\newblock \emph{Technometrics}, 34\penalty0 (1):\penalty0 1--14, 1992.

\bibitem[Lawless(2002)]{lawless2002statistical}
Jerald~F Lawless.
\newblock \emph{Statistical models and methods for lifetime data}, volume 362.
\newblock John Wiley \& Sons, New Jersey, 2002.

\bibitem[Lawless(1987)]{lawless1987regression}
Jerald~Franklin Lawless.
\newblock Regression methods for {P}oisson process data.
\newblock \emph{Journal of the American Statistical Association}, 82\penalty0
  (399):\penalty0 808--815, 1987.

\bibitem[Loeber et~al.(1998)Loeber, Farrington, and
  Waschbusch]{loeber1998serious}
Rolf Loeber, David Farrington, and Daniel Waschbusch.
\newblock \emph{Serious and {V}iolent {J}uvenile {O}ffenders}.
\newblock SAGE Publications, California, 01 1998.
\newblock ISBN 9780761920403.
\newblock \doi{10.4135/9781452243740.n2}.

\bibitem[Lorentz(1986)]{lorentz1986bernstein}
G.G. Lorentz.
\newblock \emph{Bernstein {P}olynomials}.
\newblock AMS Chelsea Publishing Series. Chelsea Publishing Company, New York,
  1986.
\newblock ISBN 9780828403238.
\newblock URL \url{https://books.google.com.br/books?id=nXiRAKzpdjUC}.

\bibitem[Meeker et~al.(2022)Meeker, Escobar, and
  Pascual]{meeker2022statistical}
William~Q Meeker, Luis~A Escobar, and Francis~G Pascual.
\newblock \emph{Statistical Methods for Reliability data}.
\newblock John Wiley \& Sons, New Jersey, 2022.

\bibitem[Migon et~al.(2014)Migon, Gamerman, and Louzada]{migon2014statistical}
Helio~S Migon, Dani Gamerman, and Francisco Louzada.
\newblock \emph{Statistical inference: an integrated approach}.
\newblock CRC press, 2014.

\bibitem[Morris et~al.(2019)Morris, Wheeler-Martin, Simpson, Mooney, Gelman,
  and DiMaggio]{morris2019bayesian}
Mitzi Morris, Katherine Wheeler-Martin, Dan Simpson, Stephen~J Mooney, Andrew
  Gelman, and Charles DiMaggio.
\newblock Bayesian hierarchical spatial models: Implementing the {B}esag {Y}ork
  {M}olli{\'e} model in stan.
\newblock \emph{Spatial and Spatio-Temporal Epidemiology}, 31:\penalty0 1--30,
  2019.

\bibitem[Osman and Ghosh(2012)]{osman2012nonparametric}
Muhtarjan Osman and Sujit~K Ghosh.
\newblock Nonparametric regression models for right-censored data using
  {B}ernstein polynomials.
\newblock \emph{Computational Statistics \& Data Analysis}, 56\penalty0
  (3):\penalty0 559--573, 2012.

\bibitem[Paternoster and Brame(1997)]{paternoster1997multiple}
Raymond Paternoster and Robert Brame.
\newblock Multiple routes to delinquency? a test of developmental and general
  theories of crime.
\newblock \emph{Criminology}, 35\penalty0 (1):\penalty0 49--84, 1997.

\bibitem[Petersilia(1980)]{petersilia1980criminal}
Joan Petersilia.
\newblock Criminal career research: A review of recent evidence.
\newblock \emph{Crime and Justice}, 2:\penalty0 321--379, 1980.

\bibitem[Piquero et~al.(2003)Piquero, Farrington, and
  Blumstein]{piquero2003criminal}
Alex~R Piquero, David~P Farrington, and Alfred Blumstein.
\newblock The criminal career paradigm.
\newblock \emph{Crime and justice}, 30:\penalty0 359--506, 2003.

\bibitem[Piquero et~al.(2007)Piquero, Farrington, and
  Blumstein]{Piquero_Farrington_Blumstein_2007}
Alex~R. Piquero, David~P. Farrington, and Alfred Blumstein.
\newblock \emph{Key Issues in Criminal Career Research: New Analyses of the
  Cambridge Study in Delinquent Development}.
\newblock Cambridge Studies in Criminology. Cambridge University Press, 2007.

\bibitem[Piquero et~al.(2012)Piquero, Jennings, and
  Barnes]{piquero2012violence}
Alex~R Piquero, Wesley~G Jennings, and JC~Barnes.
\newblock Violence in criminal careers: A review of the literature from a
  developmental life-course perspective.
\newblock \emph{Aggression and Violent Behavior}, 17\penalty0 (3):\penalty0
  171--179, 2012.

\bibitem[Prentice et~al.(1981)Prentice, Williams, and
  Peterson]{prentice1981regression}
Ross~L Prentice, Benjamin~J Williams, and Arthur~V Peterson.
\newblock On the regression analysis of multivariate failure time data.
\newblock \emph{Biometrika}, 68\penalty0 (2):\penalty0 373--379, 1981.

\bibitem[{R Core Team}(2023)]{R}
{R Core Team}.
\newblock \emph{R: A Language and Environment for Statistical Computing}.
\newblock R Foundation for Statistical Computing, https://www.R-project.org/,
  Vienna, Austria, 2023.
\newblock URL \url{https://www.R-project.org/}.

\bibitem[Ribeiro and Oliveira(2022)]{ribeiro2022}
Ludmila Ribeiro and Valéria Oliveira.
\newblock Reincidência e reentrada na prisão no {B}rasil.
\newblock \emph{Instituto Igarapé}, 57\penalty0 (1):\penalty0 219--223, 2022.

\bibitem[Rigdon and Basu(2000)]{rigdon2000statistical}
Steven~E Rigdon and Asit~P Basu.
\newblock \emph{Statistical methods for the reliability of repairable systems}.
\newblock Wiley New York, 2000.

\bibitem[Sapori et~al.(2017)Sapori, Santos, and Maas]{sapori2017fatores}
Luis~Fl{\'a}vio Sapori, Roberta~Fernandes Santos, and Lucas Wan~Der Maas.
\newblock Fatores sociais determinantes da reincid{\^e}ncia criminal no
  {B}rasil: o caso de {M}inas {G}erais.
\newblock \emph{Revista Brasileira de Ci{\^e}ncias Sociais}, 32:\penalty0
  1--18, 2017.

\bibitem[{Stan Development Team}(2021)]{rstan}
{Stan Development Team}.
\newblock {RStan}: the {R} interface to {Stan}, 2021.
\newblock URL \url{https://mc-stan.org/}.
\newblock R package version 2.21.3, https://mc-stan.org/.

\bibitem[Stollmack and Harris(1974)]{Stollmark1974}
Stephen Stollmack and Carl~M. Harris.
\newblock Failure-rate analysis applied to recidivism data.
\newblock \emph{Operations Research}, 22\penalty0 (6):\penalty0 1192--1205,
  1974.
\newblock ISSN 0030364X, 15265463.
\newblock URL \url{http://www.jstor.org/stable/169997}.

\bibitem[Tollenaar and van~der Heijden(2013)]{Heijden2013}
N.~Tollenaar and P.~G.~M. van~der Heijden.
\newblock Which method predicts recidivism best?: a comparison of statistical,
  machine learning and data mining predictive models.
\newblock \emph{Journal of the Royal Statistical Society. Series A (Statistics
  in Society)}, 176\penalty0 (2):\penalty0 565--584, 2013.
\newblock ISSN 09641998, 1467985X.
\newblock URL \url{http://www.jstor.org/stable/23355205}.

\bibitem[Vandeviver and Bernasco(2017)]{vandeviver2017geography}
Christophe Vandeviver and Wim Bernasco.
\newblock The geography of crime and crime control.
\newblock \emph{Applied geography}, 86:\penalty0 220--225, 2017.

\bibitem[Vehtari et~al.(2015)Vehtari, Simpson, Gelman, Yao, and
  Gabry]{vehtari2015pareto}
Aki Vehtari, Daniel Simpson, Andrew Gelman, Yuling Yao, and Jonah Gabry.
\newblock Pareto smoothed importance sampling.
\newblock \emph{arXiv preprint arXiv:1507.02646}, 2015.

\bibitem[Vehtari et~al.(2017)Vehtari, Gelman, and Gabry]{vehtari2017practical}
Aki Vehtari, Andrew Gelman, and Jonah Gabry.
\newblock Practical {B}ayesian model evaluation using leave-one-out
  cross-validation and {WAIC}.
\newblock \emph{Statistics and Computing}, 27\penalty0 (5):\penalty0
  1413--1432, 2017.

\bibitem[Vehtari et~al.(2021)Vehtari, Gelman, Simpson, Carpenter, and
  B{\"u}rkner]{vehtari2021rank}
Aki Vehtari, Andrew Gelman, Daniel Simpson, Bob Carpenter, and Paul-Christian
  B{\"u}rkner.
\newblock Rank-normalization, folding, and localization: An improved {R} for
  assessing convergence of {MCMC}.
\newblock \emph{Bayesian Analysis}, 16\penalty0 (2):\penalty0 667--718, 2021.

\bibitem[Watanabe and Opper(2010)]{watanabe2010asymptotic}
Sumio Watanabe and Manfred Opper.
\newblock Asymptotic equivalence of {B}ayes cross validation and widely
  applicable information criterion in singular learning theory.
\newblock \emph{Journal of Machine Learning Research}, 11\penalty0
  (12):\penalty0 3571--3594, 2010.

\end{thebibliography}

\end{document}